\documentclass[]{aa}
\usepackage{graphics}
\input psfig.sty
\begin{document}
\title{The luminosity function of cluster galaxies. II. Data reduction procedures applied to
the cluster Abell 496 $^\star$.}
\subtitle{}
\author{A. Moretti \inst{1}
    E. Molinari \inst{1}
    G. Chincarini\inst{1,2}
    S. De Grandi\inst{1}}
\institute{Osservatorio Astronomico di Brera, Via Bianchi 46, I-22055 Merate, Italy
      \and Universit\`a degli Studi di Milano, via Celoria 16, I-20133 Milano, Italy}
\date{submitted on 23-3-1999}
\titlerunning{Cluster galaxy LF. Data reduction.}
\authorrunning{Moretti et al.\ }
\thanks{Based on observation carried out at ESO La Silla, Chile.}
\offprints{moretti@merate.mi.astro.it}
\maketitle

\begin{abstract}
We initiated a large project aimed to estimate the Luminosity Function
of galaxies in clusters and to evaluate its relation to cluster morphology.
With this paper
we deem necessary to outline the general procedures of the data reduction and details
of the data analysis. The cluster sample includes the brightest southern ROSAT all-sky survey clusters
with $z < 0.1 $. These have been observed in three colours $ g,~r,~i$,  and mapped up to
a few core radii using a mosaic of CCD frames.
E/S0 galaxies in the cluster core are singled out
both by morphology (for the brightest galaxies), and by colour.
The details of the data reduction procedure are illustrated via the analysis of the
cluster Abell 496, which has been used
as a pilot cluster for the whole program. The related photometric catalogue consists
of 2355 objects. The limiting magnitudes (the reference Surface Brightness is given in
parenthesis) in the various colours are respectively
$ g(25.5) = 24.14 $, $r(25.5) = 24.46 $, $i(25.0) = 23.75$.
These correspond to the limiting absolute magnitudes -12.28, -11.96 and -12.67
($H_0$=50 km/sec/Mpc).

\end{abstract}
\section {Introduction}

In 1993/1994 we started a long-range photometry program on clusters of galaxies in order to
estimate in detail the cluster Luminosity Function (LF) and the morphology of the brightest cluster
galaxies.
Our aim was
to gain more accurate knowledge on this topic both to better understand formation and evolution,
and to improve the comparison with numerical simulations.
Straightforward scientific drivers are at the basis of this investigation:
the Luminosity Function of cluster galaxies at present time is the result of cluster initial
formation and subsequent evolution - taking into account internal phenomena and external interactions.

It is reasonable, and to some extent expected, that at formation the galaxy mass
function is a universal constant.
In this case, assuming that every evolutionary process keeps a constant M/L ratio,
it would be reasonable to
expect a universal LF, even if it cannot be excluded that evolution and richness might play a
role on
this stage.
Present day observing evidence is, however, that the mass is organised into differently shaped
and differently luminous galaxies -
the galaxy population depending strongly on the cluster density and
morphology.
It would be strange if Nature, in the unfolding of this multivariate process,
could set to work such
a fine-tuning as to maintain the exact
proportionality between mass
and luminosity, even assuming a universal initial mass function.

The assumption of a universal LF for all the clusters (Colless 1989 and,
more recently, Threntham 1997, 1998) might therefore be too coarse of a tool for
characterizing the cluster population.
Infall and ICM-galaxies interaction might further perturb the shape of the LF
during the evolution of the cluster.
In this respect it seemed of fundamental importance to evaluate the faint end of the LF.
Meanwhile, important work has been published on this topic following
the excellent papers on the Virgo Cluster by Binggeli et al. (1988).
Biviano et al. (1995) approached the study by selecting a catalogue
of bright galaxies in the Coma cluster.
Undoubtedly, this direct method is a sound way to proceed, but the construction of
a spectrophotometric catalogue of a large number of rich clusters demands an unaffordable
amount of time with a 4 meter class telescope.

Another very interesting photometric approach is that of Bernstein et al. (1995), for the
same Coma cluster, where particular attention has been given to the faint end of LF.
In that work, however, the bright part remains ill-defined.

A general consideration of the different studies is the limited application
of their results, often making it impossible to compare directly the
catalogue and Lfs. This led us to build our consistent photometric catalogues.

In this paper we outline at first the selected sample: other authors might be interested in
this bookkeeping (avoiding or comparing duplications) and it will help the reader to follow
 our work to its the completion.

Secondly, we detail our observational strategy
and methods of data reduction, particularly in those points where they differ from the
standard analysis used in the literature. They will then form a basic reference for other papers
in preparation. The observing strategies are strongly related and tuned to the
data analysis methods. These procedures have first been applied to the cluster
Abell 496 (see also Molinari et al. 1998, paper I, for discussion on LF), for which we publish
here the photometry.


\section {The Project}
\subsection {The sample}
The sample has been selected from the catalogue given in De Grandi
et al. (1999) by choosing only clusters at declination $<0^o$, with
X-ray fluxes measured in the 0.5-2.0 keV energy band larger than
$10^{-11}$ erg cm$^{-2}$ s$^{-1}$, and with extended X-ray
emission (i.e., sources with probability to be point-like, as computed
by De Grandi et al. 1997, smaller than 1\%).
The resulting sample of 20 clusters is reported in Table \ref{tab:campio}.
Columns list the main X-ray and optical properties for each source
as follows:
{\it{Column (1)  ---}} Cluster name.
{\it{Column (2)  ---}} X-ray position: J2000.0 right ascension (hh mm ss.s).
{\it{Column (3)  ---}} X-ray position: J2000.0 declination (dd mm ss.s).
{\it{Column (4)  ---}} Cluster red-shift.
{\it{Column (5)  ---}} Unabsorbed X-ray flux computed in the 0.5-2.0 keV band
                     in units of $10^{-11}$ erg cm$^{-2}$ s$^{-1}$.
{\it{Column (6)  ---}} Bautz-Morgan type.
{\it{Column (7)  ---}} Optical richness.
{\it{Column (8)  ---}} Status of observations (Obs.= observed)
\begin{table*}[htb]
\begin{center}
\caption[]{{The sample. Data relative to X-ray flux and red-shift are from
De Grandi et al. (1999). Data relative to optical richness and morphology are from
Abell et al. (1989). Data labelled with $^*$ are our estimate.} \label{tab:campio}}
\begin{tabular}{|l|ccccccc|}
\hline
Name     &$\alpha_{2000}$& $\delta_{2000}$& Redshift & Flux & B.M. & Rich.&Status   \\
\hline

A0085&           00h41m50.11s& -09d18m17.5s& 0.05560& $4.092^{+0.256}_{-0.244}$& I        &2 &Obs.\\
A0119&           00h56m11.69s& -01d14m52.5s& 0.04420& $2.406^{+0.854}_{-0.708}$& II-III   &2 &Obs.\\
A0133&           01h02m42.21s& -21d52m43.5s& 0.05660& $1.578^{+0.181}_{-0.166}$& I$^*$    &2 &Obs.\\
A3158&           03h42m53.06s& -53d37m43.0s& 0.05910& $2.250^{+0.206}_{-0.184}$& I-II     &2 &--  \\
A3186&           03h52m25.09s& -74d01m02.5s& 0.12700& $1.041^{+0.177}_{-0.157}$& I-II     &1 &Obs.  \\
EXO0422-086&     04h25m51.02s& -08d33m38.5s& 0.03971& $1.870^{+0.188}_{-0.179}$& I$^*$    &--&Obs.\\
A3266&           04h30m58.82s& -61d27m52.5s& 0.05890& $3.001^{+0.869}_{-0.717}$& I        &2 &Obs.  \\
A0496&           04h33m37.07s& -13d15m20.0s& 0.03284& $4.652^{+0.423}_{-0.389}$& I        &1 &Obs.\\
A3376&           06h01m37.77s& -40d00m31.0s& 0.04550& $1.504^{+0.519}_{-0.429}$& I        &1 &--  \\
A3391&           06h26m20.10s& -53d41m44.5s& 0.05310& $1.313^{+0.183}_{-0.152}$& I        &0 &--  \\
A3395&           06h27m38.83s& -54d26m38.5s& 0.04980& $1.122^{+0.289}_{-0.209}$& I        &3 &Obs.\\
A3667&           20h12m35.08s& -56d50m30.5s& 0.05560& $3.289^{+0.927}_{-0.668}$& II       &2 &Obs.\\
A3695&           20h34m46.86s& -35d49m07.5s& 0.08930& $1.247^{+0.315}_{-0.243}$& I        &2 &Obs.\\
A3822&           21h54m10.21s& -57d52m05.5s& 0.07590& $1.099^{+0.202}_{-0.170}$& II-III   &2 &Obs.\\
A3827&           22h01m58.85s& -59d57m37.0s& 0.09840& $1.517^{+0.173}_{-0.160}$& I        &2 &--  \\
A2420&           22h10m20.09s& -12d10m49.0s& 0.08380& $1.172^{+0.240}_{-0.199}$& I        &2 &Obs.\\
A3921&           22h50m03.61s& -64d26m30.0s& 0.09360& $1.201^{+0.259}_{-0.226}$& II       &2 &Obs.  \\
RX-J2344.2-422&  23h44m15.98s& -04d22m24.5s& 0.07860& $1.214^{+0.302}_{-0.232}$& I$^*$    &--&--  \\
A4038&           23h47m41.78s& -28d08m26.5s& 0.02920& $2.751^{+0.232}_{-0.221}$& III      &2 &Obs.\\
A4059&           23h57m00.02s& -34d45m24.5s& 0.04600& $1.974^{+0.189}_{-0.178}$& I        &1 &Obs.\\
\hline
\end{tabular}
\end{center}
\end{table*}
%

\begin{figure}[hbt]
\centerline{\psfig{figure=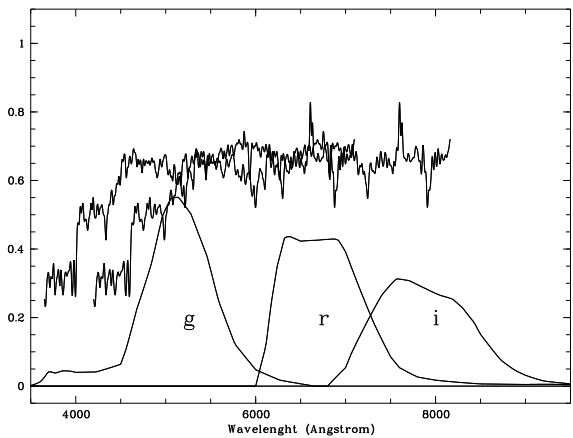,width=9cm,angle=270}}
\caption[]{The efficiency of the $g,~r,~i$ filters as function of the wavelength.
In comparison, a typical early type galaxy spectrum is superimposed
at two different red-shifts: z=0 and z=0.12, the extremes of the catalogue redshift
range.}
\label{fig:spettro}
\end{figure}
\begin{figure}[!hb]
\centerline{\psfig{figure=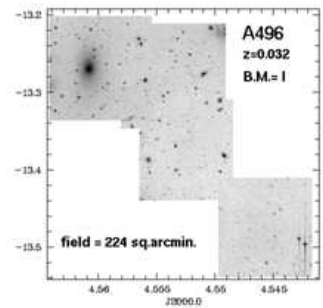,width=6cm}}
\caption[]{The observed field of the cluster Abell 496 with the cD galaxy in
the NE corner of the image. The mosaic is composed of 4 adjoining, and slightly
overlapping fields: their identification number (Table \ref{tab:journ})
increases moving from NE (field 0) to SO corner (field 3).
The angular distance between NE and SO corners is 30 arcmin.}
\label{fig:mosaic}
\end{figure}

\subsection {Imaging}
CCD observations of the sample clusters were carried out since December 1994
at La Silla with the 1.5 m Danish Telescope equipped with DFOSC camera.
For each cluster we observed a mosaic composed of 3 or 4 slightly overlapping fields
(Fig. \ref{fig:mosaic} shows the mosaic of
Abell 496). In each mosaic the centre of the first field corresponds to
the centre of the X-ray isophotes (see the Fig. 1 in Molinari et al. 1998.).
The other fields are centred along a radial direction.
For each mosaic, the typical total observed area is 250 arcmin$^2$
with a typical maximum angular distance of 30 arcmin (equal to a linear
distance of 2.5 Mpc at z=0.05).
For each cluster the observation of the most external field is used mainly to
evaluate the background.
Each field is observed with the $g,~r,~i$ filters of the Gunn photometric system
(Thuan \& Gunn 1976, Wade et al. 1979). The spectral response is illustrated in
Fig.\ref{fig:spettro} along with the observed spectrum of an elliptical galaxy.
Observations of each field consist of 3600 s exposure as a result of the integration
of 4 $\times$ 900 s different exposures.
Up to date, we have collected photometric observations of 15 out of 20 clusters
of the sample (Table \ref{tab:campio}).
 Spectroscopic observations are also being planned and will likely start shortly before
completion of the photometric sample.
This paper will deal, in particular, with the data analysis carried out for the cluster Abell 496.
However it reflects the method we will also use for the other clusters.

\section{Abell 496 image processing}
Abell 496 is a  class 1 rich cluster, Bautz Morgan type I (Abell et al. 1989),
dominated by a single central cD galaxy, MGC -02-12-039
($ \alpha_{2000} = 4^h 33' 37.7''$, $ \delta_{2000} = -13^o 15' 43.2'' $,
z=0.032).
The peak of the X-ray emission lies inside the core of the cD galaxy
(Table \ref{tab:campio}). CCD observations of the cluster were carried out during
the first observing run from 24 to 27 December 1994.
The effective field of the DFOSC camera and Thomson THX 31156 CCD is $8.68 \times 8.68$
arcmin with a single pixel corresponding to $0.508~$ arsec .
The total area of the observed field is $224~$arcmin$^2$ for each filter
(Fig.\ref{fig:mosaic}).
We list the journal of the observations in Table \ref{tab:journ}.

\begin{figure} [!htb]
\centerline{\psfig{figure=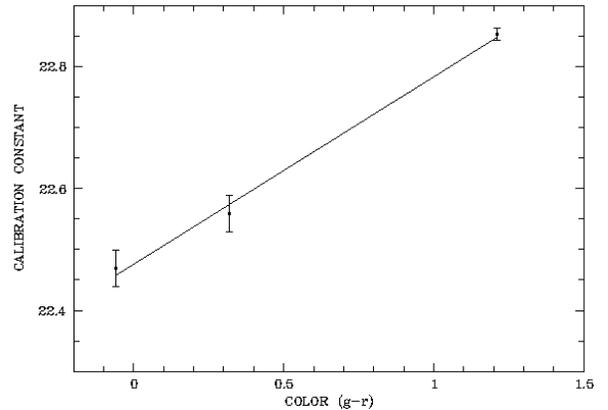,width=8cm}}
\caption[]{The calibration straight line for filter $r$. For each of the
three standard star, the typical uncertainty on the offset measure is 0.02 magnitude.
Moreover, the 3 different average offset values show a linear dependence on
the colour of the star. By the linear fit, we extrapolate the offset value
corresponding to $g-r=0$.}
\label{fig:fitis}
\end{figure}
\subsection {Flat-fielding and magnitude calibration}
Basic data reduction, including bias subtraction, flat-field correction,
magnitude calibration and cosmic rays subtraction, is done using the ESO-MIDAS
software environment.

For each filter we build two different flat-field frames.
For the first we use the dithering method
to obtain the flat field frame directly from scientific exposures
(see for example Molinari et al. 1996).
The second flat field frame is built using the median of the
distribution of the sunset and twilight sky images.
We obtain the minimum value of the ratio noise/sky, at both small and big
scales in the frames, using the first flat-fielding procedure for filter
{\it i}. For the {\it g} and {\it r} filters we adopt the average between the
two different flat-field frames, since this gives the minimum rms.
After the reduction, the typical rms of the sky is $1.5\%,~ 1\%,~ 0.75\%$ of
the background for the {\it g, r, i} frames,
respectively.
\begin{table}
\scriptsize
\begin{center}
\caption[]{The journal of observations. The Date, Universal Time,
air mass, exposure time, and seeing for each frame are shown. In each frame seeing is
calculated as the FWHM of the stars.}
\label{tab:journ}
\begin{tabular}{|l|llllll|}
\hline
Object& Filter &Date & U.T.&airmass& E.T.&Seeing \\
\hline
field0 & g  & 25-12-94 &     1:54& 1.071&900s.&1.25\\
 & & &                       2:48& 1.040&"    &1.25\\
 & & &                       3:41& 1.060&"    &1.50\\
 & & &                       4:33& 1.133&"    &1.50\\
\hline
       & r  &          &     1:37& 1.092&"    &1.25\\
&&&                          3:05& 1.041&"    &1.25\\
&&&                          3:24& 1.048&"    &1.50\\
&&&                          4:50& 1.171&"    &1.50\\
\hline
       & i  &          &     2:12& 1.054&"    &1.25\\
&&&                          2:31& 1.044&"    &1.25\\
&&&                          3:58& 1.077&"    &1.50\\
&&&                          4:16& 1.102&"    &1.50\\
\hline
\hline

field1 & g  & 26-12-94 &     1:45& 1.076&"    &1.50\\
&&&                          2:14& 1.050&"    &1.50\\
&&&                          3:41& 1.063&"    &1.15\\
&&&                          4:06& 1.093&"    &1.15\\
\hline
       & r  &          &     1:12& 1.128&"    &1.50\\
&&&                          2:32& 1.042&"    &1.50\\
&&&                          3:24& 1.050&"    &1.15\\
&&&                          4:23& 1.121&"    &1.15\\
\hline
       & i  &          &     1:28& 1.100&"    &1.50\\
&&&                          2:50& 1.039&"    &1.50\\
&&&                          3:07& 1.042&"    &1.15\\
&&&                          4:40& 1.157&"    &1.15\\
\hline
\hline
field2 & g  & 27-12-94 &     1:21& 1.105&"    &1.30\\
&&&                          5:04& 1.233&"    &1.35\\
&&&                          5:22& 1.298&"    &1.35\\
&&              28-12-94 &   1:49& 1.064&"    &1.25\\
\hline
       & r  & 27-12-94 &     1:05& 1.134&"    &1.30\\
&&&                          4:47& 1.184&"    &1.35\\
&&&                          5:39& 1.372&"    &1.35\\
&&              28-12-94 &   1:32& 1.083&"    &1.25\\
\hline
       & i  &   27-12-94 &   1:38& 1.080&"    &1.30\\
&&&                          1:56& 1.061&"    &1.30\\
&&             28-12-94 &    0:58& 1.140&"    &1.25\\
&&&                          1:16& 1.106&"    &1.25\\

\hline
field3 & g  &  28-12-94 &     2:14& 1.046&900s.&1.25\\
&&&                          4:11& 1.114&"     &1.50\\
&&&                          5:35& 1.371&"     &1.50\\
&&&                          5:52& 1.462&600s. &1.50\\
\hline
       & r  &          &     2:31& 1.040&900s. &1.25\\
&&&                          4:28& 1.147&"     &1.50\\
&&&                          5:18& 1.297&"     &1.50\\
&&&                          6:04& 1.538&600s. &1.50\\
\hline
       & i  &          &     2:48& 1.040&900s. &1.25\\
&&&                          4:45& 1.189&"     &1.50\\
&&&                          5:02& 1.240&"     &1.50\\
&&&                          6:15& 1.618&600s. &1.50\\
\hline
\end{tabular}
\end{center}
\end{table}
Cosmic rays are identified by their appearance in only one of the
dithered images.
The stars observed as standard are selected in the photometric system of
Thuan \& Gunn (1976) and are listed in Table
\ref{tab:standa}. The offset of the calibration is measured as the
difference between instrumental magnitude (as measured with the {\it g,r,i}
filters at ESO telescope) and the magnitude of the standard stars.
In spite of the fact that we evaluate a relation between
the colour of the standard star and the magnitude off-set, we decided not
to account for the colour relation due to the paucity of the data and the
possibility of systematic errors. Most importantly, we could not apply the colour
correction to galaxies which have been detected only in one or two filters (40\% of the sample).
Choosing the best compromise, we applied in all cases the magnitude correction
equivalent to $g-r = 0$. Because of this assumption, our photometric
data differ slightly from the photometric Gunn system (typically 0.1 magnitude
for an object with g-r=1 ). The colours we measure, however, match very well
the Gunn system, because the slopes of
the calibration straight lines in the three filters are similar.

\begin{table}
\scriptsize
\begin{center}
\caption[]{Standard stars used for calibration.}
\label{tab:standa}
\begin{tabular}{|l|cc|ccc|}
\hline
Name    & $\alpha_{1950}$ & $\delta_{1950}$& $g$   & $r$   & $i$  \\
\hline
HD 84937        & 09 46 12.1      & +13 59 17.0  & 8.325 & 8.383 & 8.43 \\
Ross 683        & 08 47 46.6      & +07 49 08.0  & 11.40 & 11.08 &  -   \\
BD $-15^0 6290$ &22 50 37.5& -14 31 42.0     & 10.754& 9.544 & 8.334\\
\hline
\end{tabular}
\end{center}
\end{table}

\begin{table}[hbt]
\begin{center}
\caption[]{{k correction (Buzzoni 1995) and galactic extinction (Burstein \& Heiles 1982)
values used for the E/S0 galaxies in Abell 496.}  \label{tab:correz}}
\begin{tabular}{|l|ccc|}
\hline
filter    & {\it g} &   {\it r} &  {\it i}\\
\hline
k corr.   &0.02 &0.01 &0.01\\
gal. ext. &0.07 &0.04 &0.03\\
\hline
\end{tabular}
\end{center}
\end{table}
\subsection {Object search and analysis}
Automatic object detection and magnitude evaluation have been done by
using the INVENTORY package (West \& Kruszewski 1981) implemented in the
MIDAS environment.
Galaxies of the sample span a very large range in magnitude from the
magnitude limit (mag $\sim24$, see next section) to the isophotal
magnitude (mag $\sim 13$) of the cD central galaxy.
This range corresponds to a comparable
range in the size of the galaxies. It varies from the PSF
limit ($\sim 3$ pixels) to the isophotal radius of the cD galaxy
($\sim 100$ pixels).
Because of this inherent heterogeneity, the sample is not perfectly
suitable for automatic search and analysis of the sources.
In particular, we must separate the signal of very extended objects
from the rest of the image to avoid the problem of multiple
detection. The procedure we use is composed of the following
three points.
First, we model and subtract the light of the most extended objects.
Second, we apply the INVENTORY standard research and analysis procedure to
frames in which the remaining objects are comparable in size.
Finally, we apply the INVENTORY analysis routine to the single-object images
of the modelled and rebuilt extended objects.
Here we describe only the first point of the procedure
which is the original part.
\begin{figure*} [hbt]
\begin{tabular}{cc}
{\psfig{figure=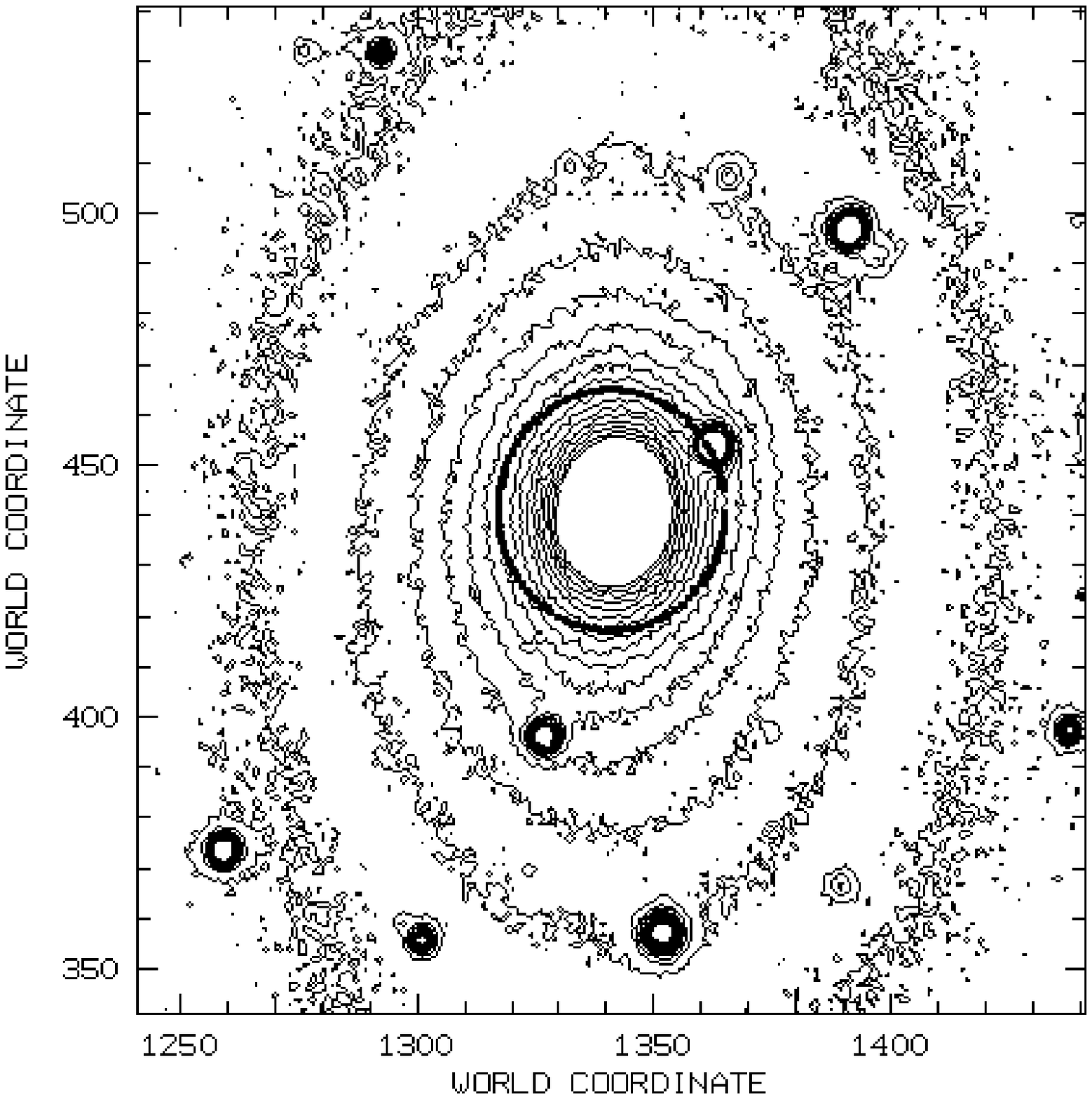,width=8cm}}&{\psfig{figure=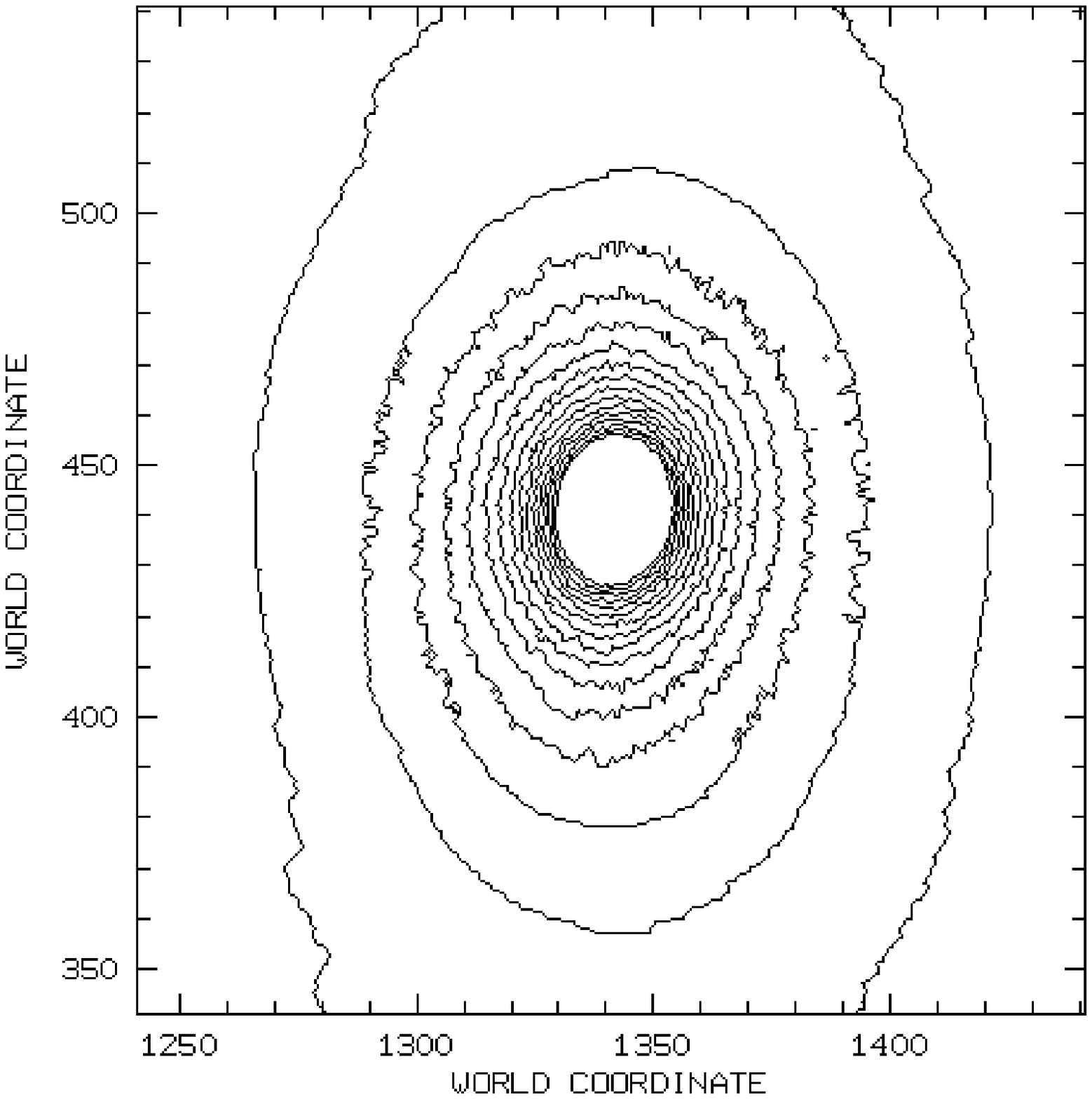,width=8cm}}\\
\end{tabular}
\caption[]{Isophotes of the cD galaxy of Abell 496 from the raw image
(left panel), and from the rebuilt model (right panel). The coordinate refer
to the pixels of the image:$ 1~pixel=0.508~$arsec. In the left panel
the circular path at 24 pixel radius is marked; the intensity profile
along this path is reported in the panel A of Fig.\ref{fig:steps}.
The model is built using raw data where possible and fit value when an
external object is superimposed on the line of sight.}
\label{fig:cdiso}
\end{figure*}
We model and rebuild the extended sources, typically giant
elliptical galaxies, with a procedure similar to the one
described by Molinari et al.(1996). We improved their algorithm by making
it more flexible.
First, for each distance from the centre of the galaxy, the algorithm analyses
the azimuthal intensity profile along the circular paths
(see the left panel in Fig. \ref{fig:cdiso}).
The projection of an elliptical isophote on the circular paths yields a
periodic variation of surface brightness, as shown in the panel A of the
Fig.\ref{fig:steps}.
It corresponds to the intensity profile along the circular path marked
on the left panel of Fig.\ref{fig:cdiso}.
The maxima correspond to the intersections of the circular path with the
major semi-axis of the isophote.
Then the algorithm fits the profile using a Fourier series and a low-pass
filter. This procedure eliminates the physical and geometrical high frequency
noise due to the discrete nature of the CCD pixel grid.
Finally, we calculate the distribution of the differences between the data and the fit:
we exclude from the profile the points whose intensity is greater than 3 times
the standard deviation of the distribution (Fig.\ref{fig:steps}, panel B).
Those points are replaced by the exact fit values. By iterating a few times
the procedure, we can separate the signals of the superimposed sources
(Fig.\ref{fig:steps} panels C), without any assumption on the shape of the isophotes.
\begin{figure} [hbt]
\centerline{\psfig{figure=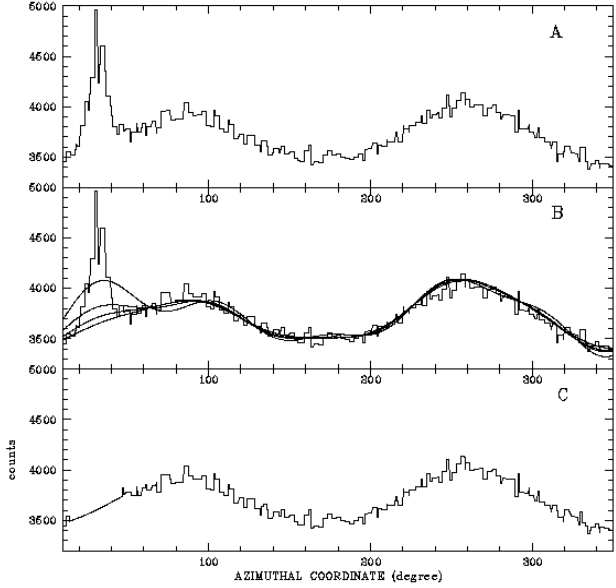,width=9cm}}
\caption[]{Steps of the modelling procedure.
A. The raw elliptical isophote is projected on a circular path.
The profile shown here corresponds to the 24 pixel radius of the
cD galaxy of Abell 496 as shown with a marked line in the left panel
of Fig.\ref{fig:cdiso}.
The azimuthal coordinate has the zero point toward
the right of the image, and it increases counterclockwise.
The periodic shape of period $\pi$ of the profile is evident:
the two maxima are at 90 and 270 degrees, corresponding to the intersections
between the path and the major axis of the elliptical isophotes
(see Fig.\ref{fig:cdiso}).
The profile of a superimposed source is evident at 30 degrees
as a departure  from the periodic shape. We can find the superimposed
object along the path marked in Fig.\ref{fig:cdiso} at 30
degrees from the 0 point of the azimuthal coordinate.
The high frequency noise in the profile shape is due both to Poissonian
and geometrical noises.
B. Fit procedure is performed repeatedly excluding step by step
the external object identified at 3 $\sigma$.
C. When an external object is identified, the extended object is
rebuilt using the fit value. Otherwise, the profile is left untouched.}
\label{fig:steps}
\end{figure}
We also made the algorithm more flexible by introducing other geometrical
parameters. In particular, we allow for the exclusion
of selected angular profiles intervals from the calculation of the Fourier
coefficient of the trigonometric series.
Intervals to be excluded are selected by visual inspection.
The exclusion option is useful when two objects of comparable size overlap
and have very close intensity maxima.
In this way, we can rebuild the hidden isophothes assuming a central
symmetry. In Fig.\ref{fig:cdiso} we compare the isophotes of the raw image of
the cD galaxy (left panel) with the rebuilt model (right panel).
The rebuilt model is then subtracted from the original frame to keep
the photometric analysis of very extended sources separated.

Although time-consuming (due to its interactiveness), this procedure yields accurate
photometric measurements of both the extended and small sources.
The described procedure, in fact, allows the complete photometric
analysis of the surface brightness of the extracted objects
(see Sect 5.1 for the Abell 496 cD).
Contrary to other popular automated programs (e.g. SExtractor, Bertin \& Arnout 1996),
we do not assign a pixel and its value to a unique object, but partition the
flux in each pixel among the different objects detected. Thus the isophotes are
recovered in their shape and intensity for all sources.
%

\section {The catalogue of Abell 496}
\subsection {Isophotal magnitude definition}
To define properly an isophotal magnitude we first need to consider some
definitions
and correlations (see also Trentham 1997).

\subsubsection{Isophotal versus total magnitude}
\begin{figure} [!htb]
\centerline{\psfig{figure=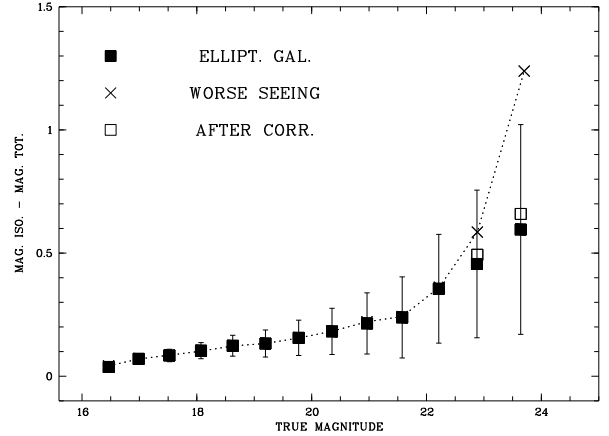,width=9cm,angle=270}}
\caption[]{The differences between the isophotal magnitude and the total magnitude of
an elliptical galaxy (seeing=1.3 arsec) are plotted (filled squares)
versus the total magnitude. Dashed line and crosses show the feature of the same elliptical
galaxy with an artificially degraded seeing (1.6 arsec).
Open squares show the seeing-degraded galaxy after the correction
performed according to the relationship seeing-threshold.}
\label{fig:magcor}
\end{figure}

The difference between total and isophotal magnitude is the difference between
the total flux, extrapolation of the curve of growth, and the flux integrated
within a
fixed SB value.
To simulate such difference, we extract from the frames some bright sources
($\sim $ magnitude 16) of different morphological types and integrate the total flux
on an extrapolated model. We then increase the magnitude up to our frame limits by
dividing the original flux by a numerical coefficient.
In this way, we obtain a list of expected total magnitudes in the range of interest.
We compare these values with the isophotal magnitudes as measured by the analysis
routine with the threshold listed below.
The amplitude of the differences is dependent on the source profile.
In our data at $r \sim 24$ the differences range from 0.1 mag for point like
sources to few tenth of mag for E0/E6 galaxies and, little more than a magnitude for
disk dominated objects (Fig.\ref{fig:magcor} shows the case of an
elliptical-r$^{1/4}$-galaxy).
The difference is seeing dependent. To show the independence
we convolve the original frames (seeing$\simeq$ 1.3 arsec) with a Gaussian point
spread function to simulate worse seeing (1.6 arsec).
The effect is illustrated in Fig.\ref{fig:magcor}.
\subsubsection{Dependence on the seeing}
\begin{figure} [!htb]
\centerline{\psfig{figure=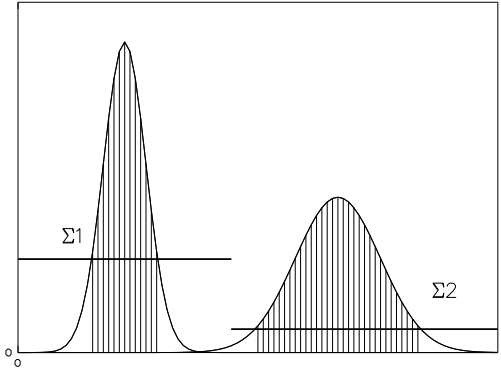,width=8cm}}
\caption[]{The two Gaussian profiles simulate the same object observed with
different PSF. The profiles are the projections of two bidimensional profiles
with the same normalization and different FWHM. The marked areas represent the
same quantity of flux.
They represent the isophotal fluxes with different thresholds at two different
seeing levels. According to equation (1), the second threshold $\Sigma_2$ is
chosen in a such a way that the isophotal flux of the left profile is kept constant.}
\label{fig:soggaus}
\end{figure}

To reach internal consistency on frames obtained with different seeing
we must correct the isophotal magnitudes for the seeing of each frame.
Our approach is as follows. We choose not to apply directly any correction to the
isophotal magnitude, but, varying the value of the SB of the last isophote
as a function of the seeing of the frame, we ensure that the isophotal magnitude
value of a fixed morphological type always corresponds to the same fraction of the
total flux of the source.
The procedure is easily justified. Consider, for simplicity, a source with a Gaussian
spatial brightness profile: in this case different seeing levels correspond to
different values of the standard deviation $\sigma$ (Fig.\ref{fig:soggaus}) and the
problem has a simple analytical solution.
Let us consider a bidimensional symmetric Gaussian profile $I_1$ with $\sigma = \sigma_1$;
given the threshold $ \Sigma_1$ we have to consider the flux $\cal F$ subtended by $I_1$
from 0 to $r_1$, such that $ I_1(r_1) = \Sigma_1 $ :
$$ {\cal F} = {1 \over 2\pi \sigma^2} \int_0^{r_1} e^{{-r^2\over 2\sigma^2}} 2 \pi r dr~.$$
After the integration, we can write it as function of $\Sigma_1$
$$ {\cal F} = 1 - 2 \pi \sigma_1^2 \Sigma_1 ~~.$$
Therefore given a different
$ \sigma = \sigma_2$ (and the same normalization), the same isophotal flux $\cal F$
is obtained using the threshold $ \Sigma_2$ such that
$$ \Sigma_2 = ({\sigma_1 \over \sigma_2})^2~\Sigma_1~.~~~~~ (1) $$
We conventionally assume a limit surface brightness value as threshold
in a frame with a certain seeing value and we use the relationship (1) to find the
correct threshold in the other frames.
The reference values of the limiting isophote SB are $ 25.5,~ 25.5,~ 25.0~ $
mag/arsec$^2$ for {\it g, r, i } filters, respectively with PSF=1.3 arsec.

\begin{figure} [!htb]
\centerline{\psfig{figure=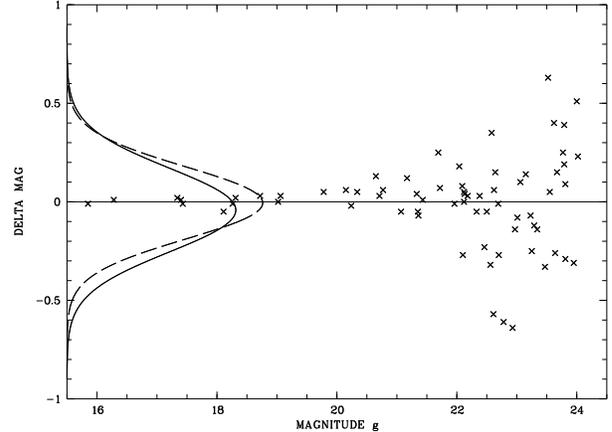,width=9cm,angle=270}}
\caption[]{The distribution of the differences between 2 measures with
different seeing ($1.3$ arcsec vs $1.4$ arcsec ) of 75 sources after the correction.
The distribution of the differences of the measures before the correction is shown
with the solid line and it is slightly asymmetric. The dashed line shows the symmetric
distribution after the correction.}
\label{fig:gaupoi}
\end{figure}

The relation (1) has been deduced in the case of Gaussian profile source.
We find that the corrections drawn from (1) give good results also for different
morphological types as shown in Figs. \ref{fig:magcor} and \ref{fig:gaupoi}.
In Fig. \ref{fig:magcor} we show, in the case of an elliptical-r$^{1/4}$ galaxy,
the difference between isophotal and true magnitude at two different
seeing levels (one artificially degraded), and the difference after the correction.
At low luminosity the correction substantially removes the seeing dependence.

The quality of the correction discussed above can be tested in the intersection
regions of two overlapping frames, which have been obtained in different seeing
conditions.
In this region we have 2 different measures performed with different seeing
of a list of sources of random magnitude and morphological type.
For the differences between the 2 independent
measures, we expect a symmetric distribution with a dispersion exponentially increasing
with the magnitude due to the Poissonian uncertainty.
If we remove this dependence by normalizing by an exponential factor, we expect a Gaussian
distribution.
In Fig. \ref{fig:gaupoi} we can observe that the distribution of the measures performed
with the same threshold is slightly asymmetric; after adopting the threshold corrected
according the relation (1) we find that the distribution of differences is perfectly
symmetric as a test of riliability of the method described.
\subsection {Sample completeness}
\begin{figure}[!hbt]
\centerline{\psfig{figure=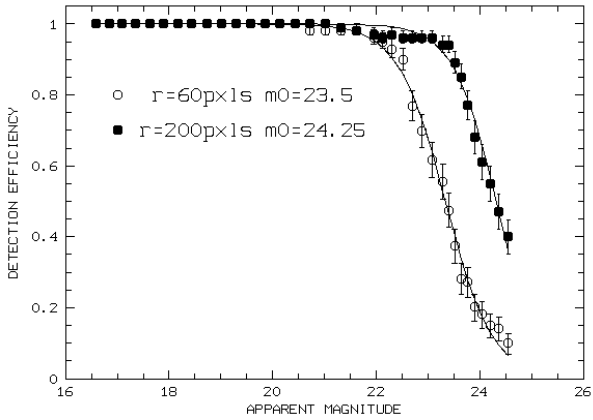,width=8cm}}
\caption[]{Bootstrap results at two different distances from the centre
of the cD galaxy are shown. The two different curves are fitted by Fermi-Dirac
function with different value of the characteristic magnitude $m_0$.
Going off centre $m_0$ increases: at fixed magnitude, finding
a faint galaxy is easier. The uncertainty of the test results is
estimate by binomial statistic and $1~\sigma$ level is shown in the
figure.}
\label{fig:fermifit}
\end{figure}

Background statistical variations and source crowding may affect
the accuracy of the automatic detection routine and the completeness of
the photometric catalogue.
We use a bootstrapping technique to test the sensitivity of our results to both
factors.
First, we extract the image of a giant elliptical galaxy from one of the frames.
Then, dividing by a numerical coefficient, we generate a set of more than
30 different images for each filter in the relevant range of isophotal magnitudes:
$ 16.07 \leq r \leq 24.56,~ 15.86 \leq g \leq 24.85,~15.87 \leq i \leq 24.01 $.
The test images are added to the observed frames positioned at
25 subsequent distances from the centre of the cluster (assumed to be in
the centre of cD galaxy).
For each value of the distance from the centre and magnitude, we repeat this
procedure 100 times in each filter, randomly changing the angular
coordinate of the added test image.
These 100 repetitions are divided in small groups in different runs
to avoid bias due to artificial additional crowding.
This allows us to estimate the probability of detecting a galaxy
of magnitude $m$ at distance $r$ from the centre of the cluster $P(r,m)$.
For each $P(r,m)$ we estimate the uncertainty by
the binomial distribution $ P_B[x,100,P(r,m)]$, which gives the probability
of observing $x$ successes on 100 attempts given a probability
$P(r,m)$ for a single success.
At a fixed distance $r$ from the centre we find a 100 \% detection rate
for bright galaxies, and a drop in the rate at characteristic magnitude
$ \sim m_0$ (Fig. \ref{fig:fermifit}).
The analytical formula of this function, given by a fit performed with
a Fermi function is:
$$  P(r,m) = {1 \over e^{{m-m_0\over c}} + 1}~. $$
We also find that $m_0$ depends on the distance r.
Smaller radii are associated with brighter $ m_0 $. The relationship can
be parameterized by an hyperbole
$$ m_0 =  m_0(r) = A-{B \over r}~~, $$
where A and B are slightly different for the 3 filters.
As we expect, this relation is affected by background statistical
variation and sources crowding.
The first steep increase of  $m_0$ is due to crowding effect of the
central part of the cluster and to the cD halo.
The flat shape near an asymptotic value is due to
the statistical variations in the background noise.
The asymptotic value of $m_0$ corresponds to  50\% detection probability independently
of any crowding effect and for each filter we assume it as the limiting
magnitude value of the catalogue (24.14, 24.46, 23.75, for filter $g,~ r,~ i$
respectively).
The test is performed on the raw image, without the exclusion of the bright,
extended objects. Indeed, we stress that subtracting the signal from
extended sources (see previous section) does not substantially improve
the automatic routine detection capability of faint galaxies.
\begin{table}[!hbt]
\begin{center}
\caption[]{{By using function $ P(r,m) $, we can estimate
our sample completeness. Here we give the completeness value of the
last three-1magnitude bin for each filter.} \label{tab:completez}}
\begin{tabular}{|l|llc|}
\hline
         & App. mag.~bin& Abs. mag.~bin& Compl. (\%)\\
\hline
filter g &[23.14,~24.14]&[-12.39,~-11.39]&  $67^{+4.5}_{-4.8}$ \\
         &[22.14,~23.14]&[-13.39,~-12.39]&  $95^{+1.7}_{-2.4}$ \\
         &[21.14,~22.14]&[-14.39,~-13.39]&  $99^{+0.9}_{-0.6}$ \\
\hline
filter r &[23.46,~24.46]&[-12.07,~-11.07]&  $68^{+4.5}_{-4.7}$ \\
         &[22.46,~23.46]&[-13.07,~-12.07]&  $96^{+1.5}_{-2.2}$ \\
         &[21.46,~22.46]&[-14.07,~-13.07]&  $99^{+0.9}_{-0.6}$ \\
\hline
filter i &[22.75,~23.75]&[-12.79,~-11.79]&  $71^{+4.3}_{-4.6}$ \\
         &[21.75,~22.75]&[-13.79,~-12.79]&  $97^{+1.2}_{-2.0}$ \\
         &[20.75,~21.75]&[-14.79,~-13.79]&  $99^{+0.9}_{-0.6}$ \\
\hline
\end{tabular}
\end{center}
\end{table}
\subsection {Stars and galaxies}
\begin{figure}[!hbt]
\centerline{\psfig{figure=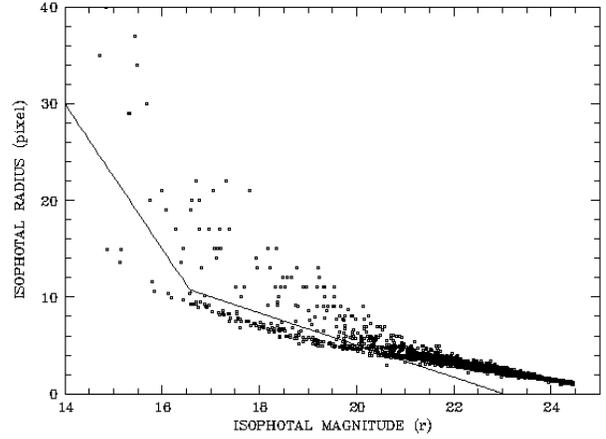,width=8cm}}
\caption[]{The isophotal magnitude-isophotal radius plane: we can easily
identify 279 bright stars up to r=20.75. At fainter magnitudes our
data do not allow us to classify morphologically the sources of our
sample. The continuous line mark the separation between the star and
the galaxy fields.}
\label{fig:stargal}
\end{figure}

We identify and remove foreground bright stars from the catalogue by using the
isophotal magnitude-isophotal radius plane (fig.\ref{fig:stargal}).
In this plane there is a clear distinction between two different
populations of sources up to the magnitude $r~\simeq$  20.75: within this range stars
have smaller isophotal radius than galaxies at any given magnitude.
We cannot classify fainter stars morphologically. Their identification
from our data can be achieved only in a statistical way by estimating
the contamination level of our sample.
In the bright part of the catalogue ($14.0 < mag.< 20.75$) we identify 279
stars. In the remaining part of the catalogue, we expect to have 290 foreground
faint stars (Table \ref{tab:statab}), about $15\%$ of the total of faint sources
(Robin et al. 1995).
The star contamination level falls under $10\%$ if we limit our analysis to
the ``sequence'' galaxies sample (see next section).
\begin{table} [!hbt]
\begin{center}
\caption[]{{Number of faint, unclassified, stars expected within our
catalogue, divided into 1 magnitude bins. First row refers to the whole
sample, second row refers to ``sequence'' colours. The last column
reports ratios between stars and galaxies:
contamination level of the whole sample is about $15\%$, whereas
sequence galaxies contamination is under $10\%$.} \label{tab:statab}}
\begin{tabular}{|l|cccc|c|}
\hline
MAG  & 21.25 & 22.25 & 23.25 & 24.25&  TOTAL   \\
\hline
stars&    56 & 75    &   94  &  65  & 290/1867 \\
stars&    13 & 7     &   13  &  14  &  47/530  \\
\hline
\end{tabular}
\end{center}
\end{table}
\subsection {Error estimate}
The Poissonian uncertainty is the largest source of error in our photometric
measurements and can be estimated by comparing independent magnitude
measurements of the same objects.
In our sample, we have independent photometric measurements of the objects
belonging to the intersections of two adjoining fields.
As shown in Table \ref{tab:nogg}, they represent a statistically significant
subsample.

\begin{table}[hbt]
\caption[]{{Number of objects belonging to the intersection of different fields.}
\label{tab:nogg}}
\begin{center}
\begin{tabular}{|l|ccc|}
\hline
Filter                   & g &  r & i  \\
\hline
$Field~1 \cap Field~0$   & 91& 141& 128\\
$Field~1 \cap Field~2$   & 95& 126& 107\\
\hline
\end{tabular}
\end{center}
\end{table}

\begin{table*}
\begin{center}
\caption[]{A subsample of the photometric catalogue
(http://www.merate.mi.astro.it/$\sim$molinari/A496-cat.dat) reporting the 40
 brightest objects in the complete list is presented. The luminosity sorting
 has been made in the $r$ filter. The ID number refers to the position
in the whole catalogue.
}
\label{tab:40bri}
\begin{tabular}{|r|rr|ccc|ccc|cc|l|}
\hline
ID & $x$ & $y$ & $g$ & $r$ & $i$ & $Rg$ & $Rr$ & $Ri$ & $g$-$r$ & $g$-$i$ & note \\
\hline

1429 & -652.6 & 419.8 & 15.22 & 14.70 & 14.54 & 32.0 & 35.0 & 31.0 & 0.50 & 0.68 & star \\ 
661 & -553.3 & 23.8 & 15.41 & 14.84 & 14.71 & 36.0 & 40.0 & 34.0 & 0.61 & 0.76 & star \\ 
2251 & -856.6 & 939.4 & 14.81 & 14.87 & 14.87 & 13.5 & 14.9 & 12.2 & -0.17 & -0.10 &  \\ 
1525 & -292.9 & 501.1 & 15.79 & 15.13 & 14.95 & 11.6 & 13.6 & 12.3 & 0.66 & 0.83 &  \\ 
1061 & -640.4 & 193.7 & 15.66 & 15.15 & 15.25 & 12.0 & 15.0 & 11.7 & 0.47 & 0.33 &  \\ 
748 & -272.2 & 55.5 & 15.83 & 15.30 & 15.17 & 26.0 & 29.0 & 25.0 & 0.52 & 0.65 & star \\ 
249 & -240.6 & -146.6 & 15.89 & 15.34 & 15.29 & 26.0 & 29.0 & 25.0 & 0.56 & 0.60 & star \\ 
1448 & -286.8 & 440.7 & 15.99 & 15.44 & 15.32 & 31.0 & 37.0 & 30.0 & 0.50 & 0.68 & star \\ 
968 & -631.8 & 152.6 & 15.98 & 15.48 & 15.42 & 32.0 & 34.0 & 29.0 & 0.50 & 0.58 & star \\ 
114 & -587.8 & -198.7 & 16.15 & 15.69 & 15.55 & 27.0 & 30.0 & 27.0 & 0.43 & 0.61 & star \\ 
935 & 47.3 & 138.5 & 16.27 & 15.76 & 15.59 & 18.0 & 20.0 & 19.0 & 0.54 & 0.71 & star \\ 
1543 & -437.4 & 514.5 & 16.11 & 15.80 & 15.80 & 11.0 & 11.6 & 9.9 & 0.33 & 0.33 &  \\ 
573 & -149.6 & -22.6 & 16.19 & 15.83 & 15.81 & 10.1 & 10.6 & 10.3 & 0.43 & 0.43 &  \\ 
1365 & -418.3 & 373.6 & 16.58 & 15.99 & 15.87 & 20.0 & 21.0 & 18.0 & 0.64 & 0.79 & star \\ 
113 & 159.7 & -199.1 & 16.61 & 16.08 & 15.89 & 17.0 & 19.0 & 18.0 & 0.55 & 0.74 & star \\ 
2257 & -1077.9 & 942.2 & 16.32 & 16.12 & 16.18 & 9.9 & 10.4 & 9.0 & 0.19 & 0.16 &  \\ 
355 & -61.8 & -107.4 & 16.50 & 16.18 & 16.18 & 9.7 & 10.0 & 9.2 & 0.35 & 0.32 &  \\ 
387 & -25.6 & -93.0 & 16.81 & 16.27 & 16.10 & 14.0 & 17.0 & 15.0 & 0.55 & 0.73 & star \\ 
632 & 56.8 & 8.0 & 16.91 & 16.38 & 16.19 & 12.2 & 13.6 & 12.9 & 0.56 & 0.74 & star \\ 
71 & -186.1 & -217.7 & 17.81 & 16.42 & 15.35 & 7.4 & 9.7 & 11.8 & 1.43 & 2.47 &  \\ 
480 & -25.9 & -60.0 & 16.92 & 16.44 & 16.18 & 12.0 & 15.0 & 14.0 & 0.46 & 0.72 & star \\ 
1919 & -922.3 & 681.2 & 17.49 & 16.56 & 16.28 & 8.1 & 10.4 & 9.5 & 0.91 & 1.24 &  \\ 
1200 & -339.8 & 251.2 & 17.12 & 16.58 & 16.52 & 17.0 & 19.0 & 16.0 & 0.53 & 0.62 & star \\ 
1280 & -482.8 & 304.7 & 17.74 & 16.59 & 15.89 & 7.8 & 9.3 & 9.6 & 1.18 & 1.89 &  \\ 
1522 & -633.1 & 496.8 & 16.96 & 16.61 & 16.58 & 20.0 & 20.0 & 17.0 & 0.41 & 0.50 & star \\ 
1006 & -77.7 & 175.2 & 16.81 & 16.62 & 16.65 & 9.0 & 9.3 & 8.8 & 0.25 & 0.19 &  \\ 
2032 & -1078.6 & 771.6 & 17.44 & 16.63 & 16.44 & 7.8 & 9.3 & 8.5 & 0.79 & 1.01 &  \\ 
781 & -236.8 & 69.3 & 16.96 & 16.69 & 16.72 & 8.8 & 9.3 & 8.4 & 0.22 & 0.19 &  \\ 
2308 & -908.9 & 987.2 & 17.83 & 16.70 & 15.97 & 7.1 & 9.5 & 9.6 & 1.10 & 1.87 &  \\ 
1511 & -420.9 & 489.4 & 17.24 & 16.70 & 16.61 & 20.0 & 22.0 & 19.0 & 0.52 & 0.67 & star \\ 
982 & -135.0 & 159.3 & 17.88 & 16.74 & 16.06 & 7.1 & 9.0 & 9.9 & 1.15 & 1.81 &  \\ 
1082 & -629.4 & 200.5 & 17.12 & 16.76 & 16.60 & 19.0 & 20.0 & 18.0 & 0.37 & 0.51 & star \\ 
925 & 131.5 & 132.0 & 17.31 & 16.76 & 16.61 & 15.0 & 17.0 & 16.0 & 0.55 & 0.70 & star \\ 
901 & -516.1 & 121.2 & 17.14 & 16.78 & 16.77 & 8.7 & 9.5 & 8.5 & 0.34 & 0.34 &  \\ 
826 & -67.3 & 90.6 & 17.30 & 16.80 & 16.62 & 12.0 & 13.0 & 13.0 & 0.50 & 0.68 & star \\ 
981 & -360.6 & 159.0 & 17.24 & 16.86 & 16.85 & 8.7 & 9.3 & 8.2 & 0.32 & 0.33 &  \\ 
1604 & -1083.8 & 545.4 & 17.86 & 16.87 & 16.52 & 7.7 & 10.1 & 8.9 & 0.95 & 1.34 &  \\ 
130 & -68.4 & -189.4 & 18.00 & 16.90 & 16.32 & 6.9 & 9.0 & 9.7 & 1.09 & 1.67 &  \\ 
2082 & -897.2 & 812.4 & 17.13 & 16.93 & 16.96 & 8.6 & 8.9 & 7.6 & 0.17 & 0.22 &  \\ 
478 & -580.7 & -60.5 & 17.40 & 16.97 & 16.95 & 8.2 & 9.2 & 7.9 & 0.39 & 0.41 &  \\ 

\hline
\end{tabular}
\end{center}

\end{table*}

Starting from the Poissonian statistics, due to the errors propagation
law, for the magnitude uncertainty, we expect
$$ \sigma(m)= {\rm const} 10^{0.2m}~~,$$
where the constant is given by the characteristics of the electronics.
At magnitude 22 we estimate that the uncertainty of the photometric measures
$\sigma_{22}$ is 0.20, 0.19, 0.22 magnitude for filter $g,~r,~i$
respectively.
Then, for each magnitude $m$ we can draw $\sigma_{m}$ as
$$ \sigma(m) = \sigma_{22}~10^{0.2(m-22)}~~,$$
which is the uncertainty of our photometric measures as function of
the magnitude.
\begin{figure} [!hbt]
\centerline{\psfig{figure=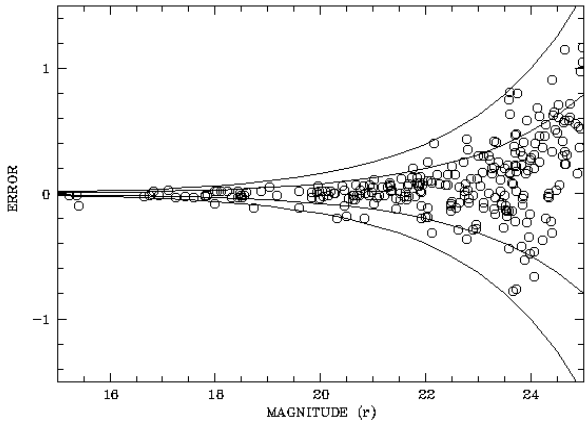,width=8cm}}
\caption[]{Differences between $r$ magnitude measurements of the same objects performed
from two frames. Plotted against magnitude, they show the expected exponential
slope. 1 and 2 $~\sigma_{m}$ level curves are shown.}
\label{fig:poiss}
\end{figure}
\begin{figure} [!hbt]
\centerline{\psfig{figure=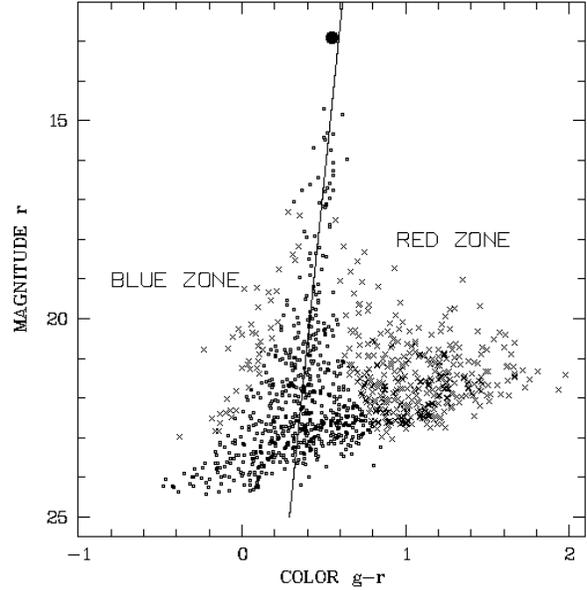,width=8cm}} \caption[]{We
split up the plane in three different regions on the basis of the
Colour-Magnitude Relation. Using Metcalfe et al.(1994) terminology
the three regions are defined as the ``blue'', ``the sequence''
and the ``red''. We identify 637 galaxies within the sequence zone
(little squares), 371 in the red zone and 47 in the blue one. The
filled circle at $r\sim13$ refers to cD galaxy isophotal magnitude
and core index colour (see next subsection).} \label{fig:cmrel}
\end{figure}

 \subsection {Description of the catalogue}
The derived catalogue consists of 2355 objects: 2076 are classified as
galaxies, 956 galaxies have magnitude measures in all three filters,
1081, 2055, 1500 galaxies have magnitude values below the
filter {\it g, r, i} limit, respectively.
We estimate {\it g-r} colours of 1058 galaxies and {\it g-i} of
955 galaxies.
The whole catalogue is available in electronic form
(http://www.merate.mi.astro.it/$\sim$molinari/A496-cat.dat),
while in Table \ref{tab:40bri} the list of the 40 brightest galaxies is
reported, and in Table \ref{tab:sintesi} we summarise the statistics of the catalogue of galaxies.
In the different
columns we list:
\begin{itemize}
\item
(1) ID number of the object;
\item
(2) EAST and SOUTH coordinates, in arseconds, with respect to the
centre of the cD galaxy;
\item
(3) {\it g, r, i} isophotal magnitudes, each computed down to the threshold
quoted in the previous section;
\item
(4) {\it g, r, i} isophotal radius;
\item
(5) {\it g-r} and {\it g-i} colours index computed through a 1.5, 3 or 5 pixel
aperture photometry, depending on the computed {\it r} isophotal radius;
\item
(6) classification of the object as star or galaxy.
\end{itemize}
\begin{table} [!hbt]
\caption[]{{Statistics of the catalogue of galaxies;
279 classified bright stars are included in the catalogue, but
not in the present summary table.
At faint magnitudes ($>20.75$) we expect $ 15\%$ of the entries
are foreground stars. In parentheses absolute magnitude limits are
reported assuming $H_0$=50 km/sec/Mpc.} \label{tab:sintesi}}
\begin{center}
\begin{tabular}{|l|cc|c|}
\hline
Skill &  Min. & Max & Gal. enters \\
\hline
$\delta_{2000}$ & $-13^o 32' 24''$ & $ -13^o 11' 26'' $ & 2077   \\
$\alpha_{2000}$ & $  4^h 32' 48''$ & $   4^h 33' 49'' $ & 2077   \\
mag. g          & 12.64 (-22.93)   & 24.14 (-12.28)     & 1081   \\
mag. r          & 12.04 (-22.62)   & 24.47 (-11.96)     & 2055   \\
mag. i          & 11.88 (-22.55)   & 23.75 (-12.67)     & 1500   \\
col. g-r        & -0.50            & 1.98               & 1059   \\
col. g-i        & -0.61            & 3.28               &  956   \\
\hline
\end{tabular}
\end{center}
\end{table}
\section{Abell 496 photometric properties}
In this section we analyse the general properties of the cluster.
We examine closely the following points. First, by means of the
Colour-Magnitude Relation, we select the main, early type,
component of the cluster population. Second, we estimate the
projected spatial distribution of the different types of galaxies
and we measure the core radius of the cluster as tracked by bright
galaxies. Third, we analyze the photometric properties of the cD
central galaxy. Fourth, we study the distribution of galaxy colour
as function of their position within the cluster core.
\subsection{The colours of the galaxies}
On the $ r/(g-r) $ plane (Fig. \ref{fig:cmrel}) we emphasize
the narrow sequence of the linear Colour-Magnitude Relation ($CMR$):
the sequence defines the locus of early type galaxies of the cluster
within the plane (Visvanathan \& Sandage,1977, Arimoto \& Yoshii,1987).
The continuous line is determined by fitting the locus of points as defined by
elliptical galaxies brighter than magnitude 18, excluding the cD galaxy.
The equation derived by the best fit
$$ CMR(r) = -0.025~r + 0.914 $$
has been extrapolated to the limiting magnitude of the frame.
The slope of the $CMR$ is consistent with that estimated by Visvanathan \& Sandage (1977)
for the Virgo cluster (see their table 1 and Fig. 1 and 2) and very similar to the
estimates given by Garilli et al. (1996). The cD galaxy fits quite
nicely the locus of the elliptical galaxies and the $CMR$ relation.
\begin{figure*} [hbt]
\begin{tabular}{|cc|}
\hline
{\psfig{figure=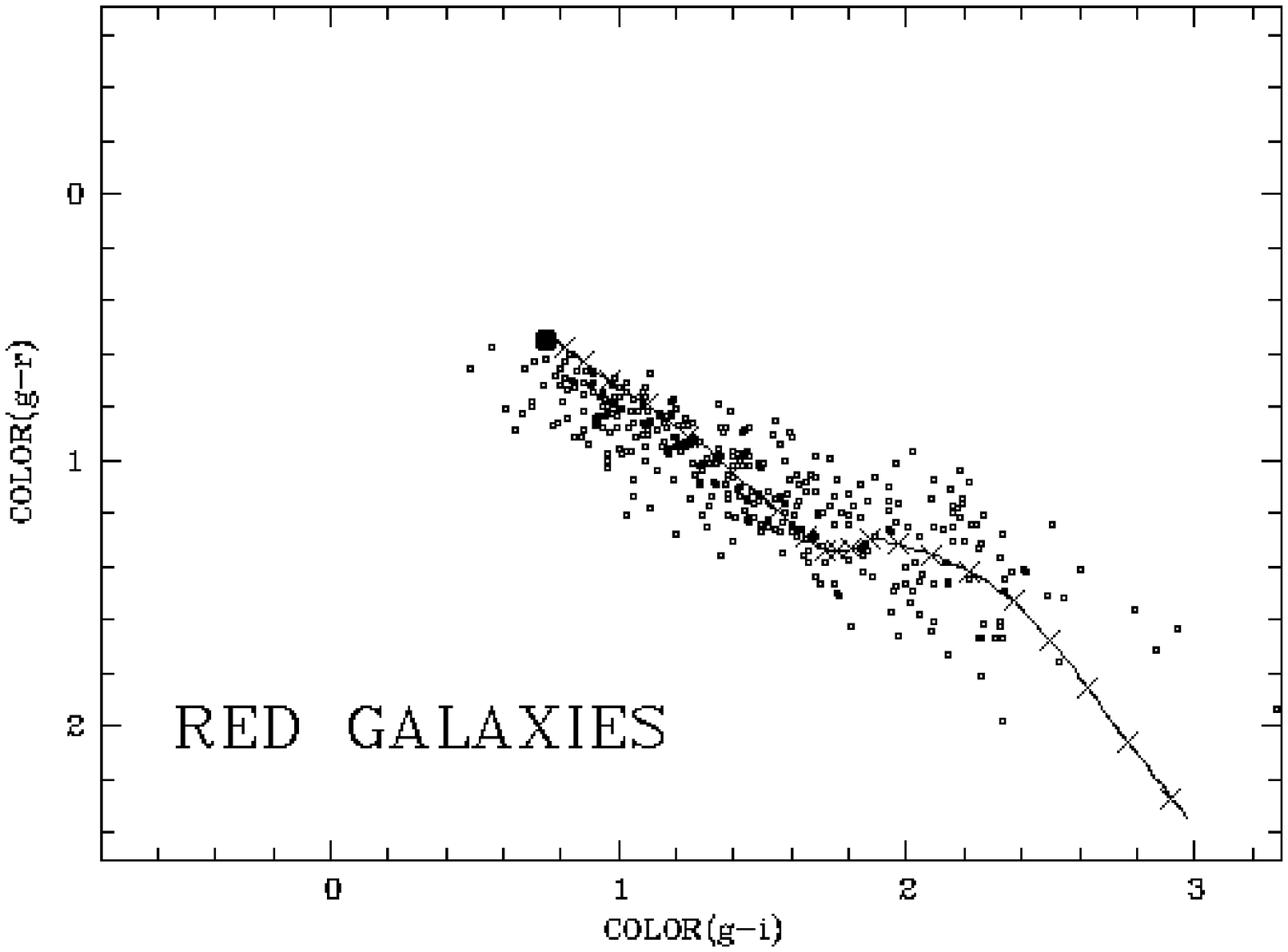,width=8cm}}&{\psfig{figure=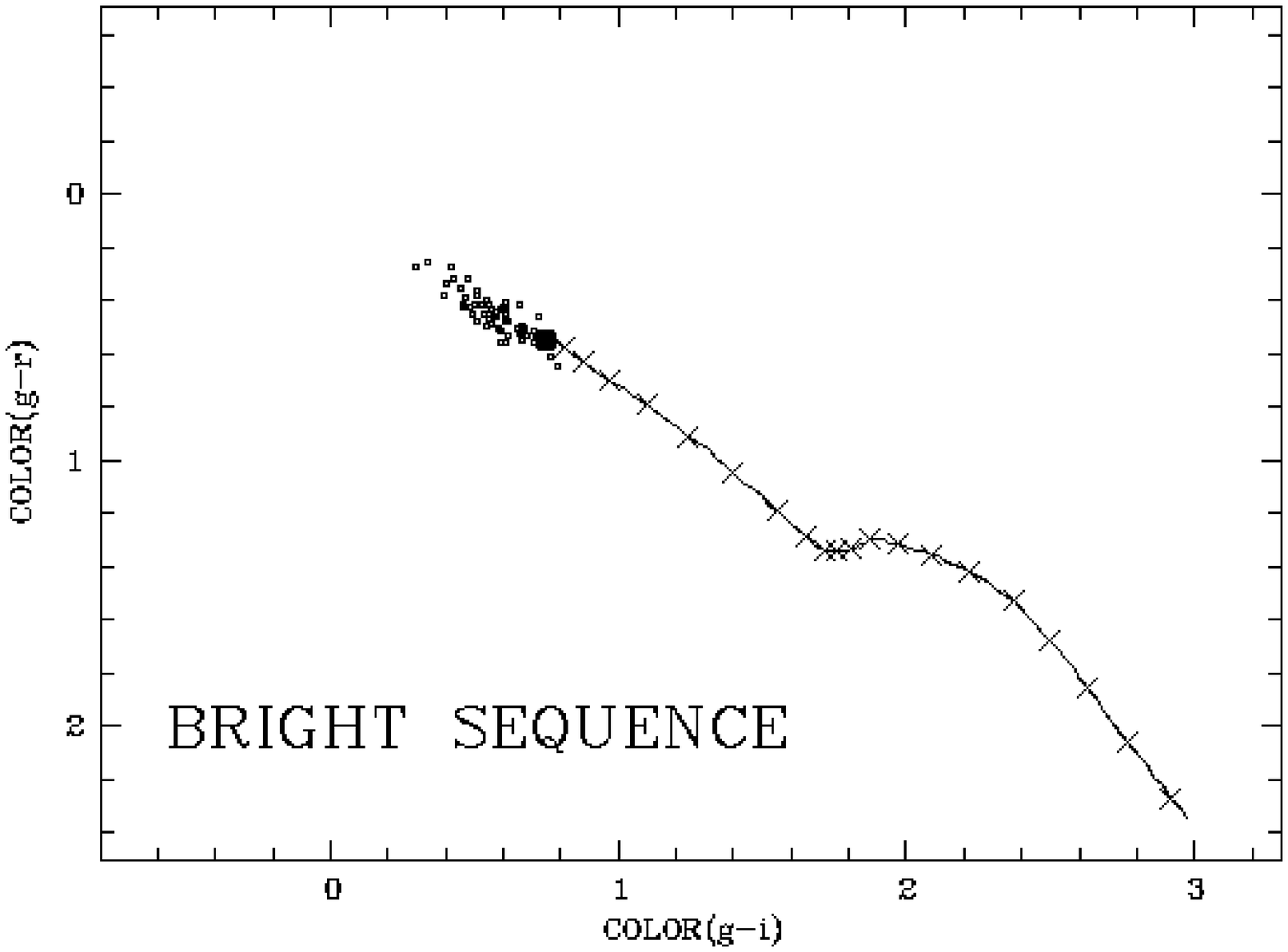,width=8cm}}\\
{\psfig{figure=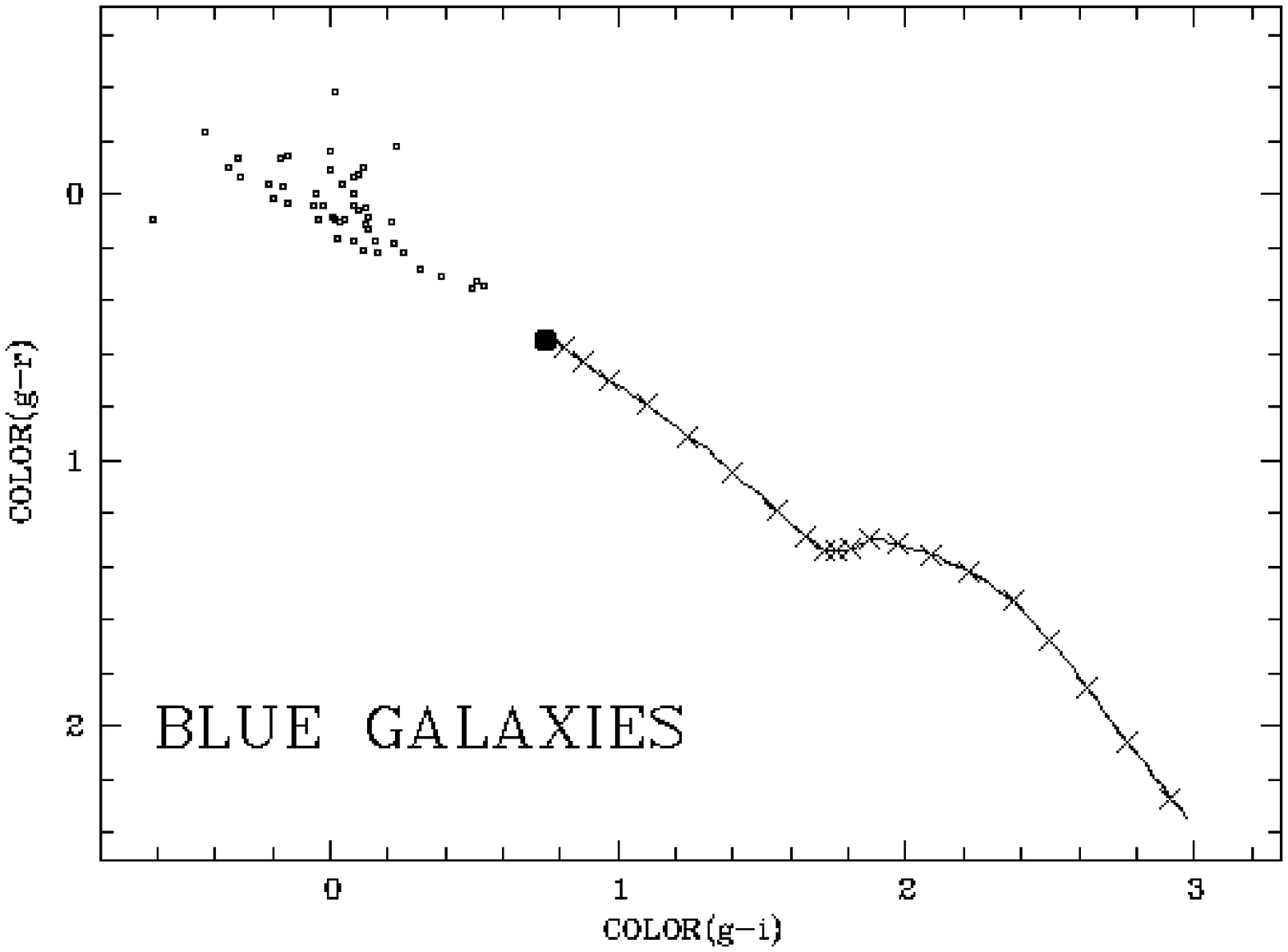,width=8cm}}&{\psfig{figure=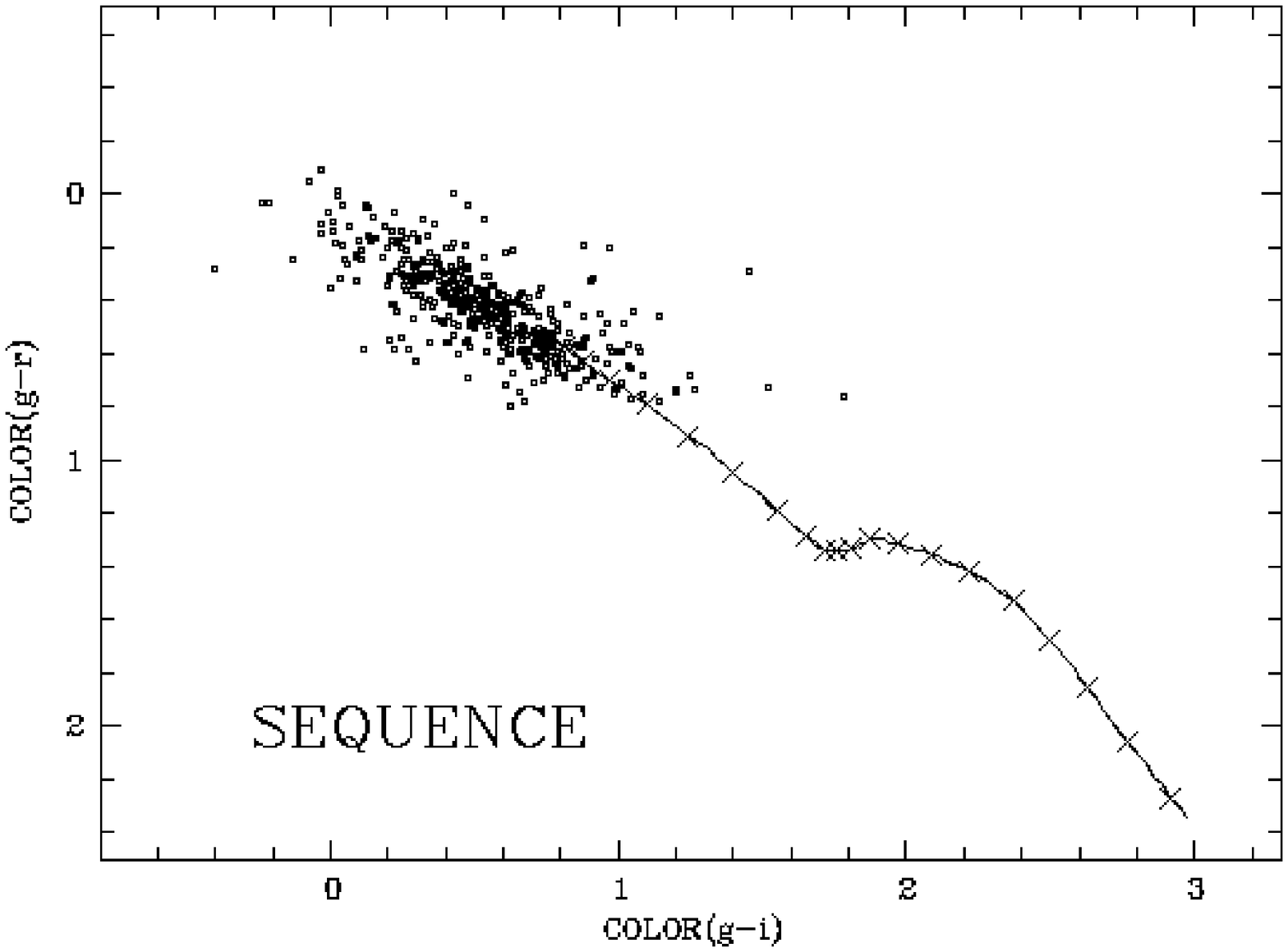,width=8cm}}\\
\hline
\end{tabular}
\caption[]{Colour-colour planes. Our data are superimposed on the
expected colours of elliptical galaxies at different red-shifts: each
cross on the continue line represents a 0.05 red-shift variation. The filled
square represents the cD galaxy, perfectly placed on the theoretical
path at red-shift 0.03. Redder galaxies show expected colours of elliptical
galaxies at higher red-shift. Sequence galaxies are slightly bluer than cD
galaxy with dispersion increasing with the magnitude
(see fig.\ref{fig:cmrel})
Finally, blue galaxies have colours unmatchable with the early type galaxy
colours.}
\label{fig:colors}
\end{figure*}
\begin{figure*} [hbt]
\begin{tabular}{|cc|}
\hline
{\psfig{figure=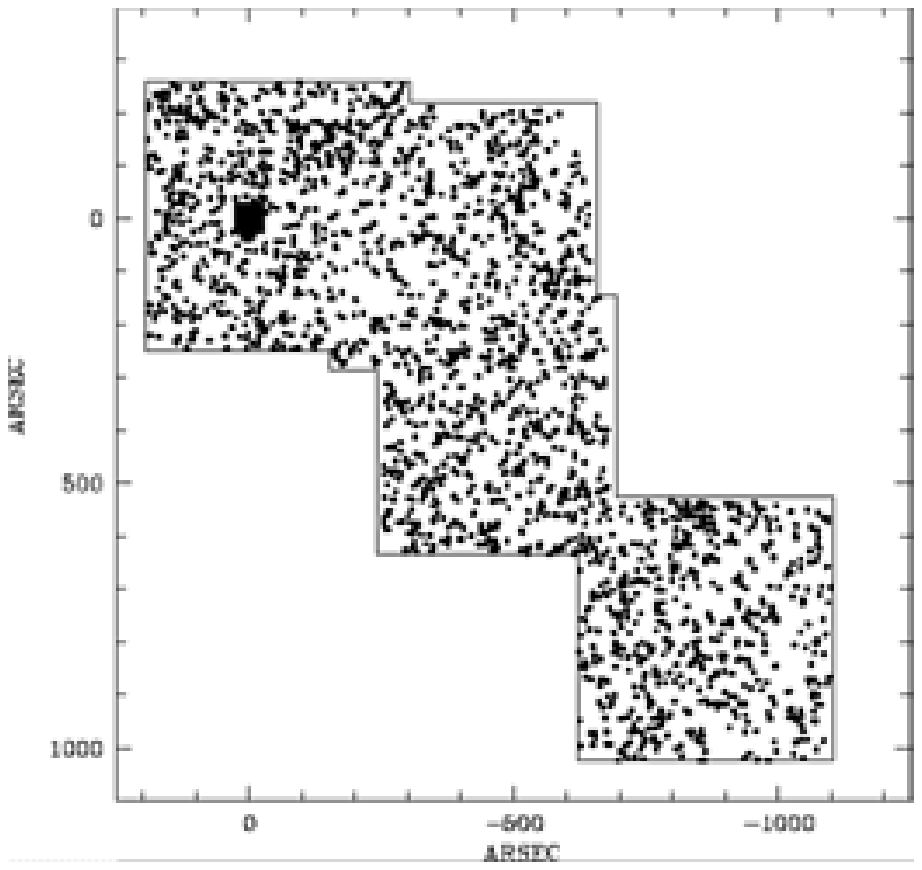,width=6cm,angle=270}}&{\psfig{figure=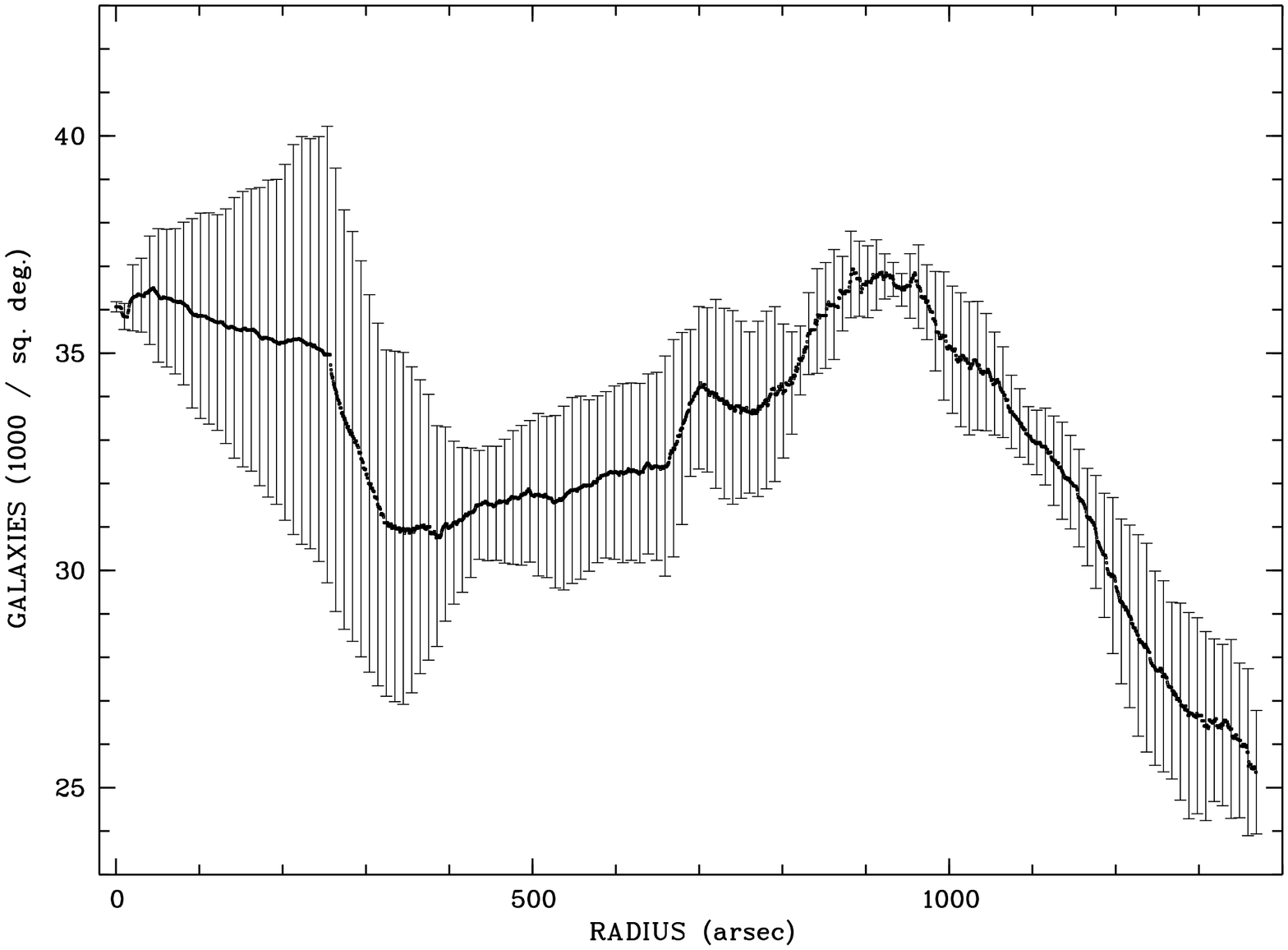,width=6cm,angle=270}}\\
{\psfig{figure=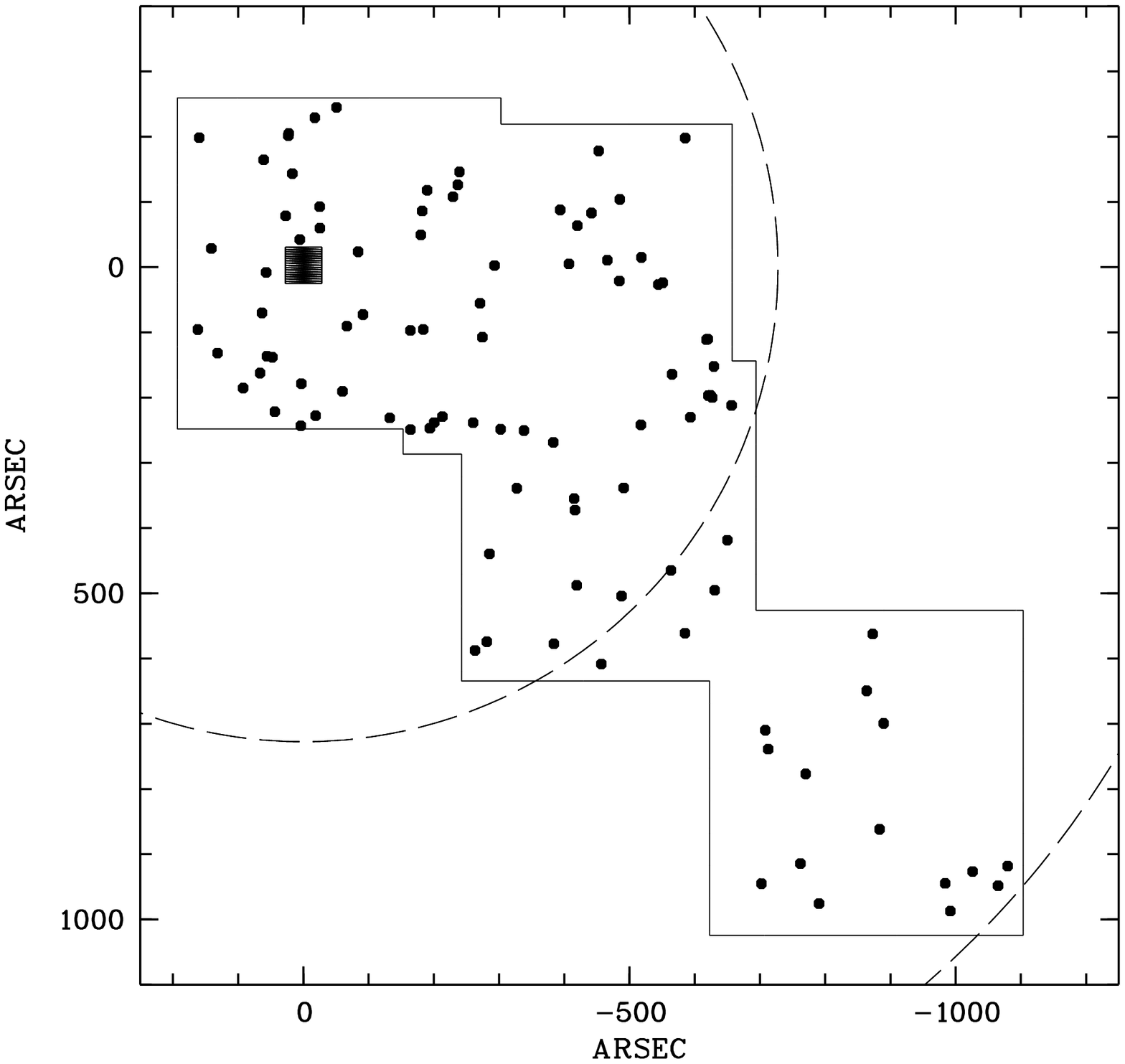,width=6cm,angle=270}}&{\psfig{figure=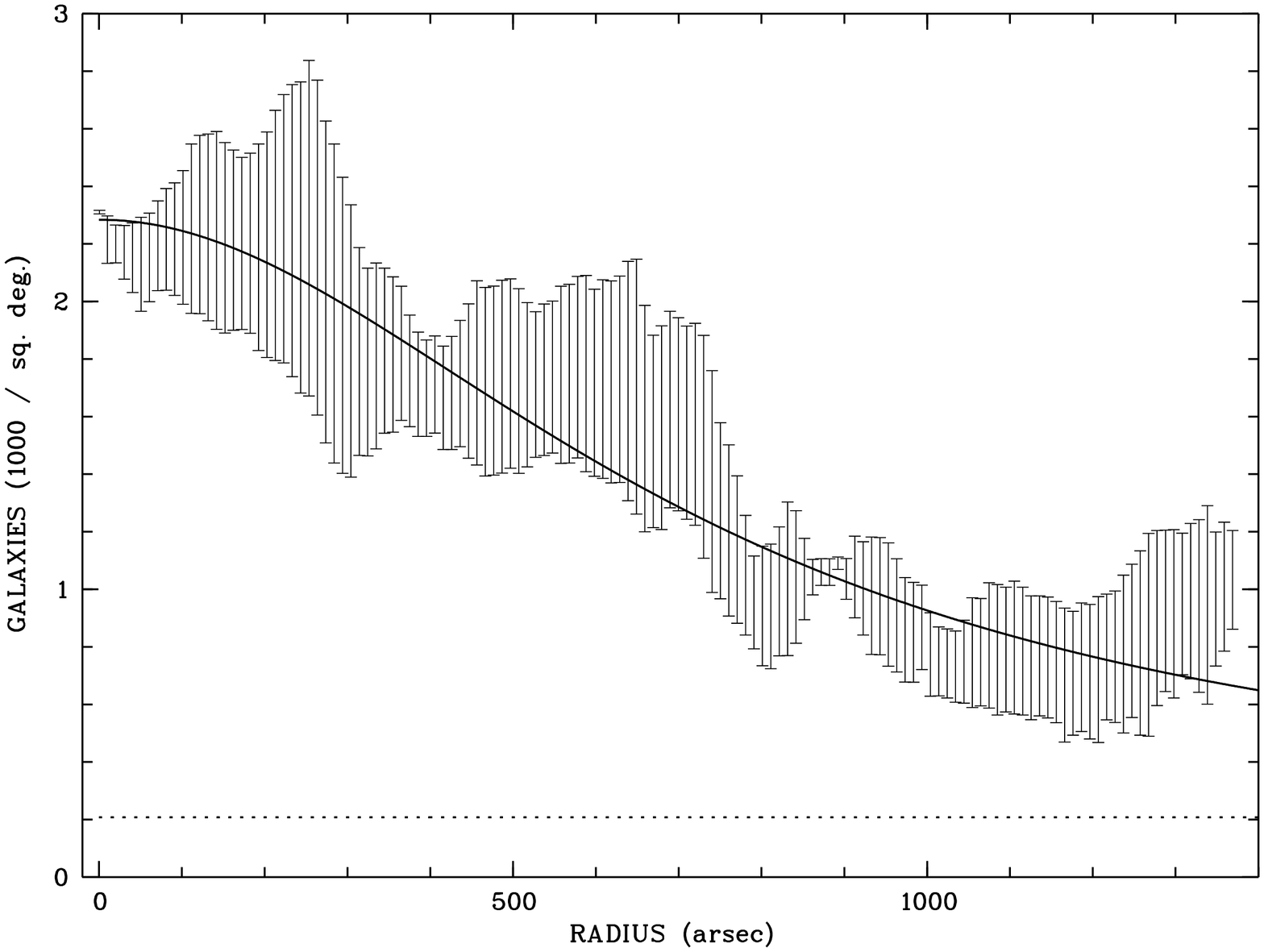,width=6cm,angle=270}}\\
{\psfig{figure=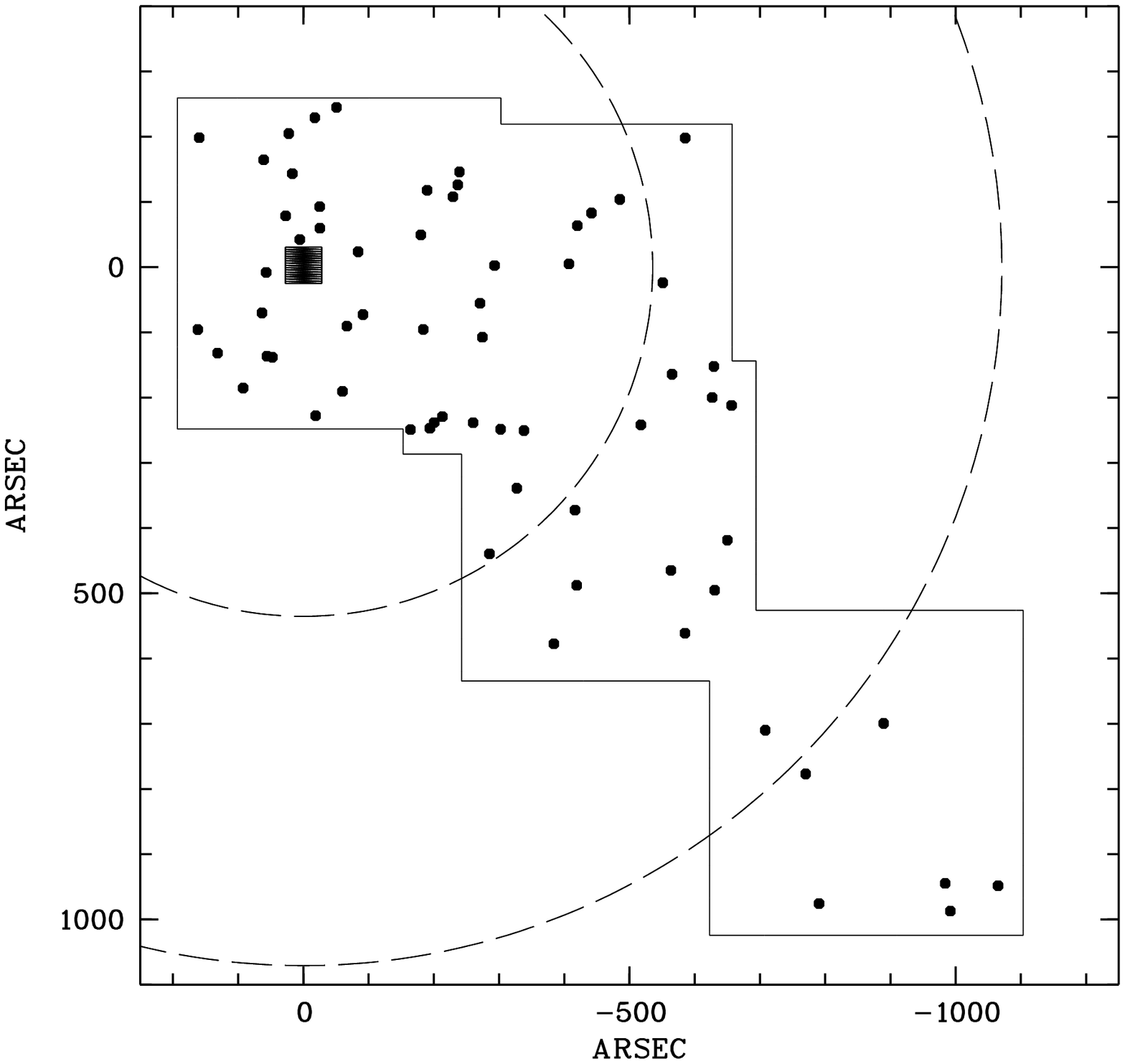,width=6cm,angle=270}}&{\psfig{figure=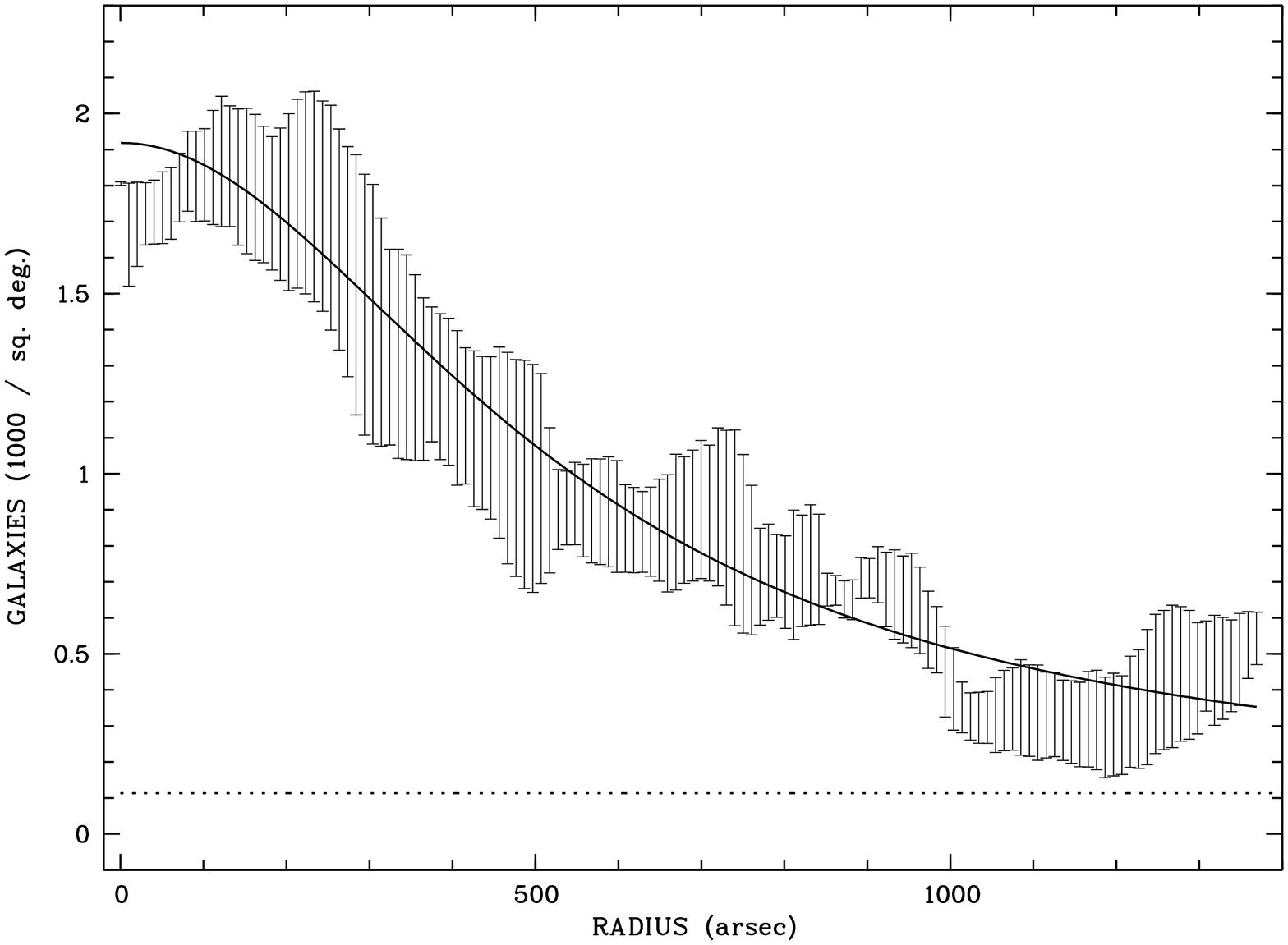,width=6cm,angle=270}}\\
\hline
\end{tabular}
\caption[]{Projected spatial distribution of all (upper panel),
bright ($ r \le 20.0$) galaxies (central panel), and bright
sequence galaxies (lower panel). We fit bright galaxies
distributions with King functions and we show the two different
core radius best values. Comparison between the upper and central
panel suggests a luminosity segregation effect; comparison between
central and lower panel suggest a colour segregation effect. The
density profile is obtained as the average of 36 profiles, and the
errors correspond to 1 standard deviation of the 36 values
distribution. In the left panel the dotted lines show 1 e 2 core
radius.} \label{fig:totell}
\end{figure*}
\begin{figure*} [hbt]
\begin{tabular}{|cc|}
\hline
{\psfig{figure=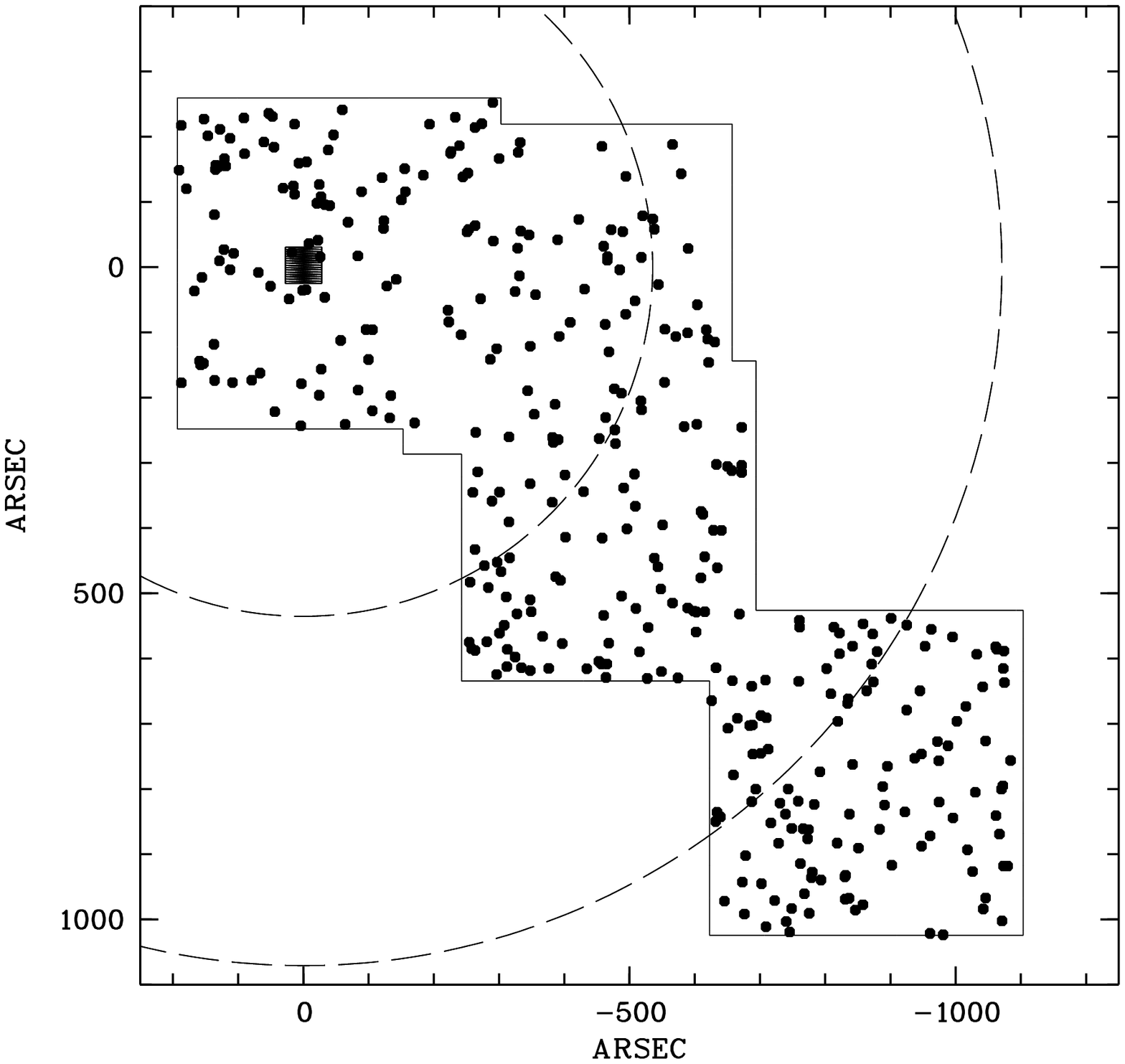,width=8cm,angle=270}}&{\psfig{figure=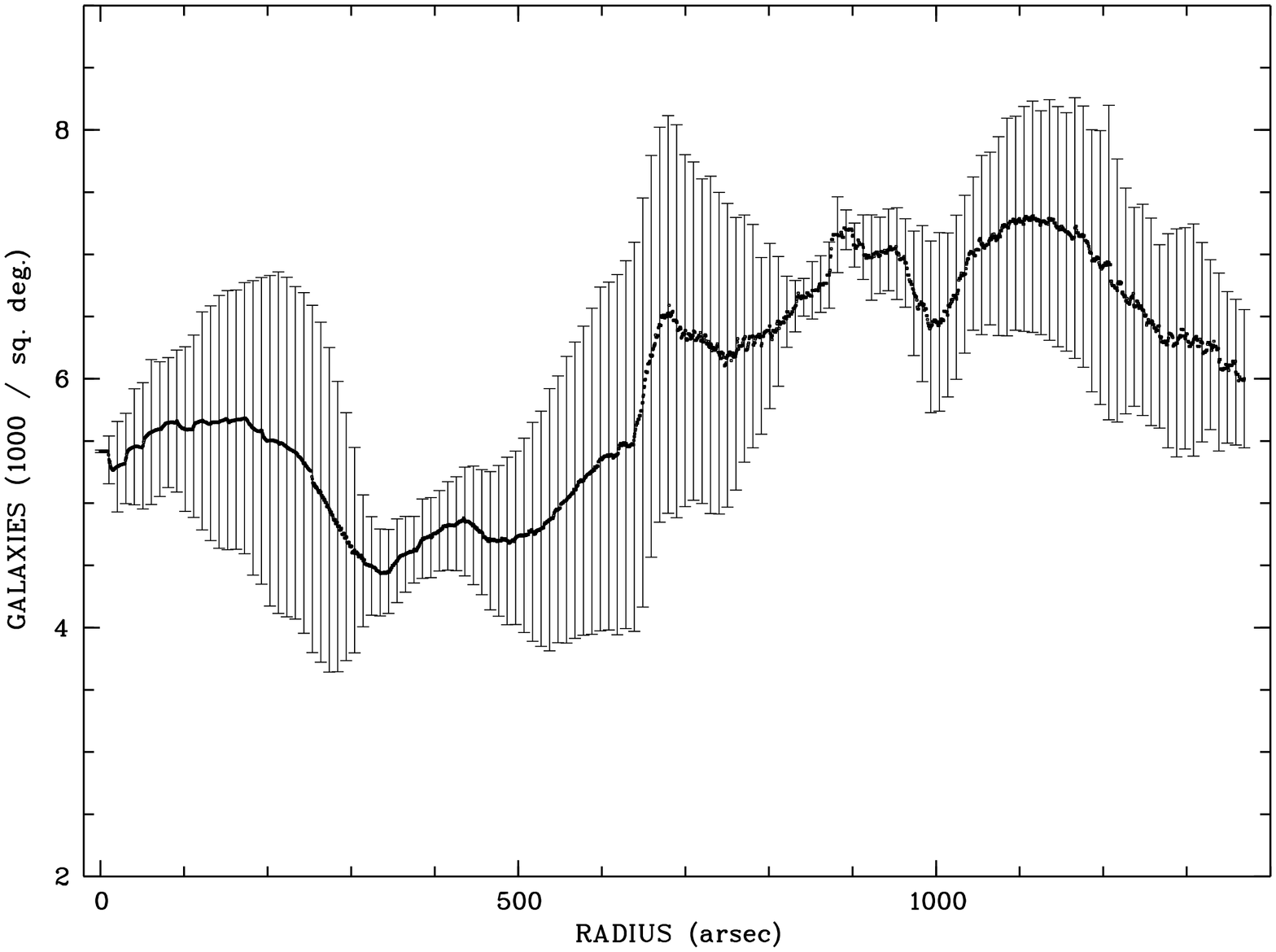,width=8cm,angle=270}}\\
{\psfig{figure=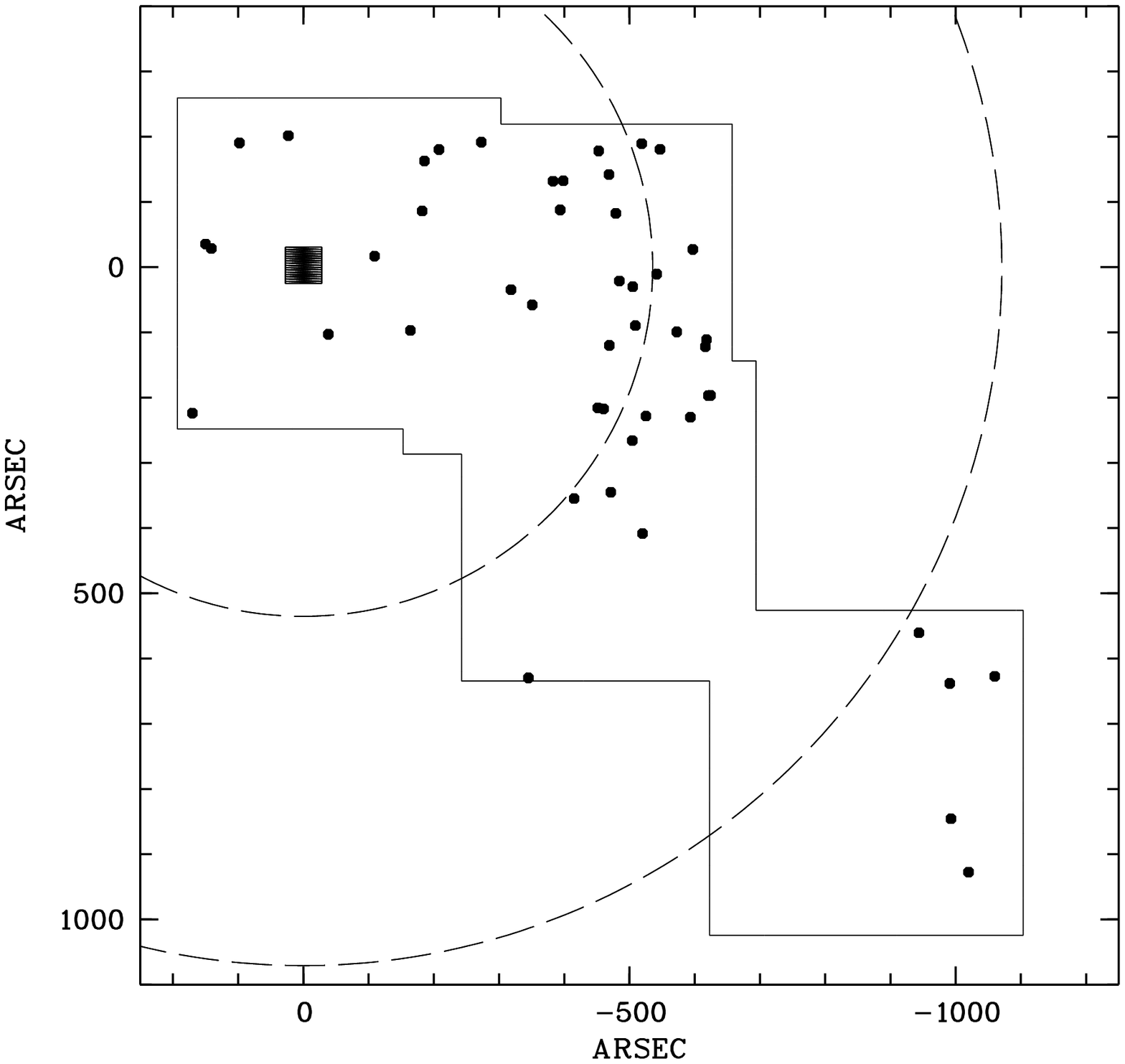,width=8cm,angle=270}}&{\psfig{figure=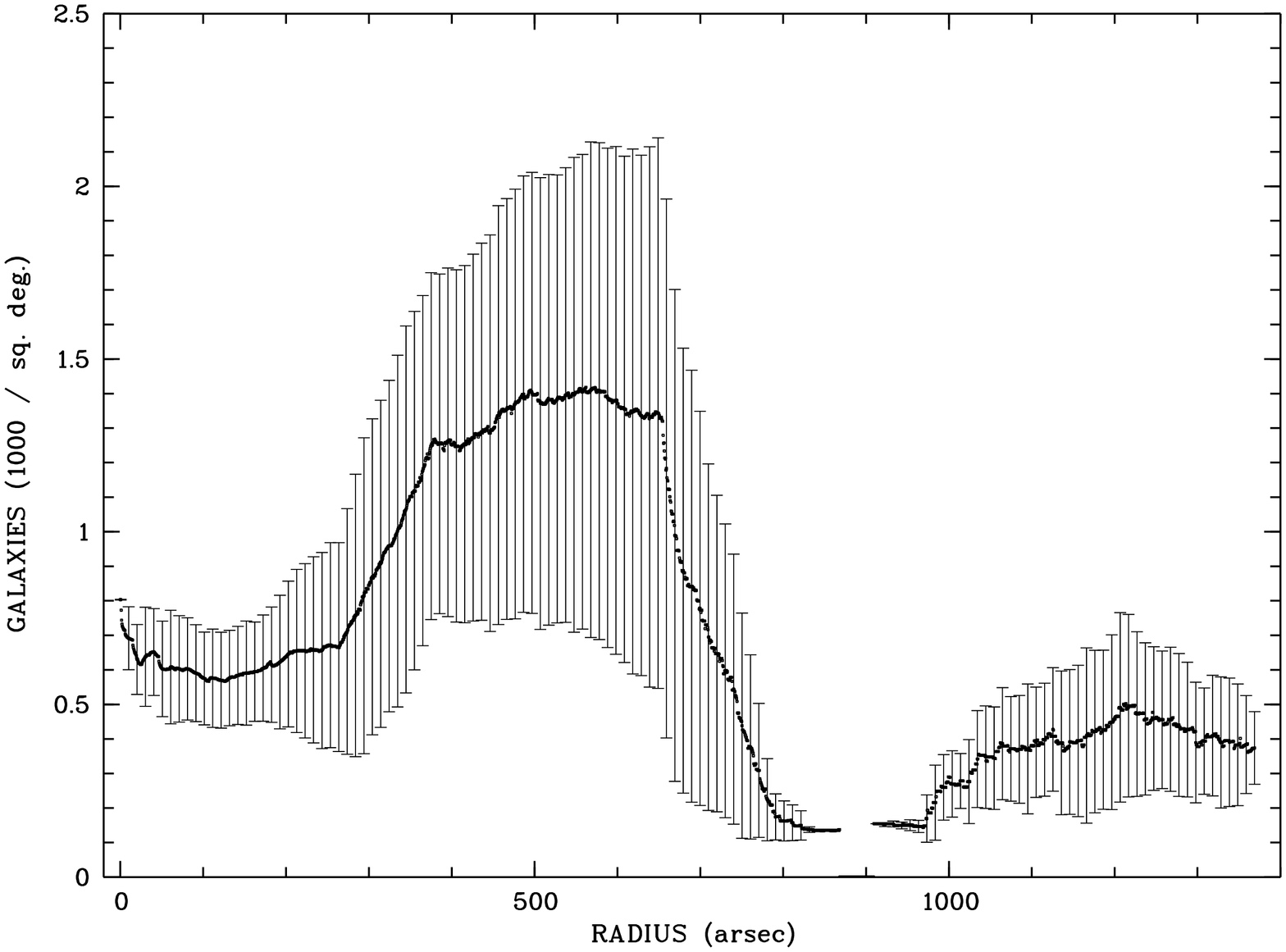,width=8cm,angle=270}}\\
\hline
\end{tabular}
\caption[]{Spatial distribution of red galaxies (upper panel) and
blue galaxies (lower panel) as labelled on the magnitude-colour
plane. Red galaxies do not show any particular behaviour linked to
cluster structure. Blue galaxies remarkably crowd at 1 core radius
distance from the centre of the cluster.} \label{fig:rosblu}
\end{figure*}

Several authors have used the $CMR$ to define cluster members (Metcalfe et al. 1994;
Biviano et al., 1995; Secker, 1996; Lopez-Cruz et al. 1997; De Propris \& Pritchet, 1998;
Molinari \& Smareglia, 1998) since, by so doing, the contamination, due to the background
galaxies, is largely reduced.
Given the analytical formula of the linear relation $CMR(r)$, determined above, we
define the ``sequence zone'' as the colour-magnitude plane region inside the curves
$$ (g-r)(r) = CMR(r) \pm (\sqrt{\sigma_g(g)^2+\sigma_r(r)^2}+0.06)~,$$
where we take into account photometric uncertainty at $1\sigma$ level
(see section 4.4) and the inherent dispersion of the relation (estimated
upon the most luminous galaxies).
The plane redward of the sequence (red zone) is expected to be mainly populated
by higher red-shift galaxies, while the blueward zone is likely the locus of cluster
and foreground late-type galaxies.

To further clarify this concept of likely membership we plot our data in the colour-colour
plane, $g-r$ versus $g-i$ (Fig.\ref{fig:colors}). The continuous line in the plane represents
the locus of points defined by elliptical galaxies at different redshifts
according to the models of Buzzoni et al.(1993). These plots are consistent with the previous
discussion: a) the cD galaxy, filled square, is near the expected location of an E galaxy at
the cluster redshift, b) galaxies located in the red zone of Fig.\ref{fig:colors} are displayed
along the sequence of higher redshift ellipticals, and c) blue galaxies do not
match the redshift sequence for elliptical galaxies.

\subsection{Spatial distribution}
The strategy we adopt for the observations has the advantage of
allowing measuring fields at a rather large distance, about 2700
pixels ($\sim 1275$ kpc) from the cluster centre in a reasonable
amount of telescope time. On the other hand we are forced to
select an {\it ad hoc\/} radial direction. That is we are more
sensitive to cluster and background field density fluctuations. We
proceed as follows. First, we build the density frame relative to
the whole mosaic. Then we divide the density frame in 36 circular
sectors centred on the cluster centre and average the contribution
of each segment at fixed radius going from the centre to the
external limit of the mosaic. The whole sample mean radial surface
density profile (Fig.\ref{fig:totell}, upper panel) does not
clearly make evident the excess of galaxies defining the cluster.

Due to the segregation effect of the most luminous galaxies, $r < 20.0$, a King profile well fits
 the density profile
at these magnitudes (Fig.\ref{fig:totell} central panel, and
Table \ref{tab:bing}). The sequence galaxies as defined by the $CMR$, with $r < 20.0$,
present a higher central concentration as indicated by the smaller core radius
(Table \ref{tab:bing}). This is also to be expected in a relaxed cluster since the $CMR$
sequence has been defined by using the bright elliptical cluster galaxies.

\begin{table}
\caption[] {{Best fit values of King function for the distribution of bright galaxies
($r<20$).} \label{tab:bing}}
\begin{center}
\begin{tabular}{|l|ccc|}
\hline
Sample & $\sigma_0~(10^3/sq.^2)$&$R_c~(arsec)$&$\sigma_{\infty}~(10^3/sq.^2)$\\
\hline
ALL      & $2.077 \pm 0.2 $ & $ 727_{-38}^{+33}$ & $0.22 \pm 0.02 $ \\
SEQUENCE & $1.807 \pm 0.2 $ & $ 497_{-25}^{+22}$ & $0.12 \pm 0.01 $ \\
\hline
\end{tabular}
\end{center}
\end{table}
Galaxies belonging to the red region of colour-magnitude plane are identified
as galaxies at higher red-shift (see Fig.\ref{fig:cmrel} and
\ref{fig:colors}). Their distribution is homogeneous
over the observed field without any link to cluster structure
(Fig.\ref{fig:rosblu} upper panel). Galaxies belonging to the
blue zone of the colour-magnitude plane are identified as cluster or foreground
late type galaxies. Their projected distribution seems to be influenced by
cluster potential: their density abruptly peaks at 1 core radius distance
from the cluster centre. This effect has been noticed also in some of the other clusters
that we are analysing.

\subsection {The central cD galaxy}
\begin{figure}[hbt]
\centerline{\psfig{figure=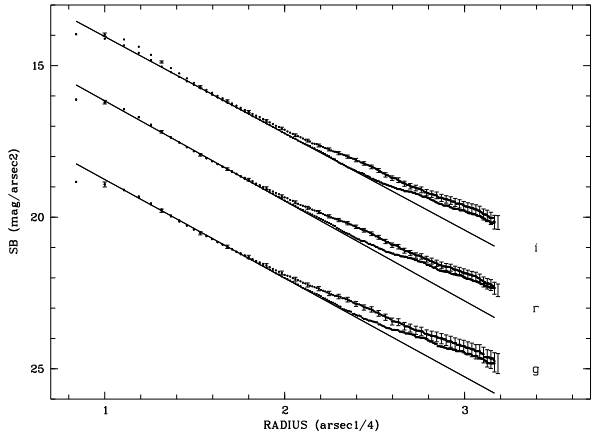,width=9cm,angle=270}}
\caption[]{The intensity profiles of the cD galaxy of Abell 496
along the N and S major semi-axis are superimposed (the $r$ and
$i$ profiles are shifted of 2 and 4 magnitude to make the figure
clearer). The excess of intensity of the northern semi-axis is
noticeable in the interval (25,50) arsec from the centre. The
straight lines represents the de Vaucouleur profile.}
\label{fig:cdgal}
\end{figure}

\begin{figure}[hbt]
\centerline{\psfig{figure=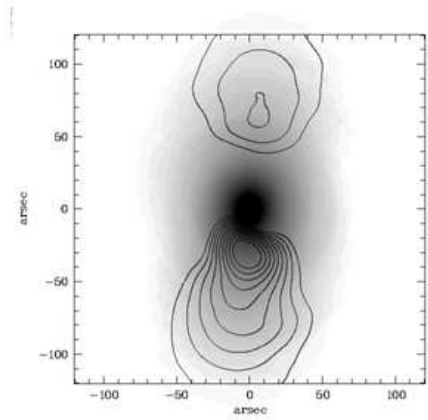,width=6cm}}
\caption[]{The filter {\it r} halo isophotes are superimposed to the image of the cD galaxy
(the North is toward the bottom of the image), the last isophote corresponding to the SB
threshold. The asymmetry of the halo emission is clearly evident.}
\label{fig:cdhal}
\end{figure}

\begin{figure}[hbt]
\centerline{\psfig{figure=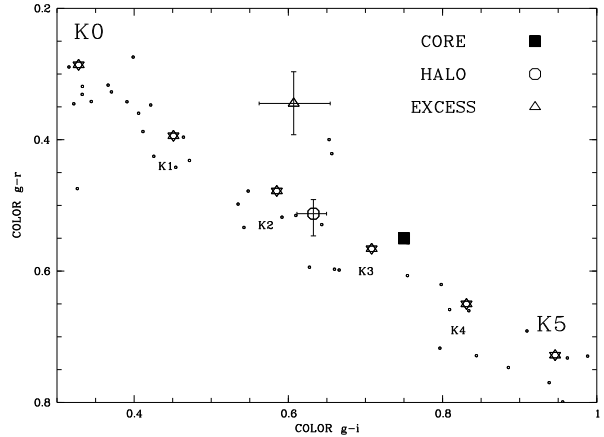,width=9cm,angle=270}}
\caption[]{The average colour index of the three components of the
galaxy is shown. They are compared with the expected colours of
the stars convoluted from the spectral catalogue of Vilnius,
Strajzhis \& Sviderskene (1972) (stars are labelled with the name
of spectral class) and also with the colours of the stars of our
catalogue (small points).} \label{fig:cdcolor}
\end{figure}

The cD central galaxy is the brightest member of the cluster: it is
2 magnitudes brighter than the second member.
In Molinari et al. (1998) its luminosity is regarded as too bright to be consistent
with other ellipticals and it is not included in the computation of LF.
However, as seen in the previous subsection, the cD magnitude and colour are
consistent with the $CMR$ extrapolated from the population of the bright ellipticals
galaxies.

cD galaxies are generally characterised by a surface brightness (SB) profile that falls off more slowly
with radius than most elliptical galaxies. In Fig. \ref{fig:cdgal} the profile of the
Abell 496 cD galaxy along the major axis is shown up to a distance of 100 arcsec
($\sim 92$ kpc) from the centre.
In this profile the presence of the halo is particularly noticeable, it departs strongly
from a de Vaucouleur law (the straight line in the figure).
The comparison of the SB profile along the northern major semi-axis (N) with
the one along the southern semi-axis (S) (Fig. \ref{fig:cdgal}) shows an evident asymmetry.
The N region of the halo exhibits an excess of intensity with respect to the S in each
of the 3 filters in the interval 25-50 arcsec of distance from the centre .
This effect is clearly depicted by the isophotes in Fig. \ref{fig:cdhal}.
In spite of the large extension of the halo, this is somewhat fainter than the core.
After fitting the core by a de Vaucouleur law, we could subtract it from the cD image and
estimate the magnitude of the halo. The derived total magnitudes in the three filters are
listed in Table \ref{tab:phopar}. As already stated, the luminosity of the core is dominant.
The average colour index of the total profile presents a gradient
toward the blue moving from the core to the outermost part of the
galaxy. This is due to the colour of the halo that is bluer than
that of the core. Within the halo itself a difference exists
between the colour of the northern hemisphere of higher surface
brightness, and the colour of the southern hemisphere. The
northern zone is bluer (marked as colour excess in Fig.
\ref{fig:cdcolor}). In other cD galaxies (see for instance
Molinari et al. 1994) the halo has been found redder than the
core. Therefore, the characteristics of the halo population are
undoubtedly related to the specific history of the cD under
consideration.

\begin{table*}
\caption[] {{Photometric parameters of the cD galaxy.} \label{tab:phopar}}
\begin{center}
\begin{tabular}{|l|ccccc|}
\hline
F & $r_e$(arsc)  &  $\mu_e$ (mag/arsc$^2$)  & $m_{tot}$  & $m_{core}$ & $m_{halo}$ \\
\hline
g &$58.3 \pm 7.5$ &$25.81 \pm 0.12$  &$12.64 \pm 0.03$ &$ 12.85 \pm 0.04$ &$14.56 \pm 0.07$\\
r &$51.7 \pm 5.1$ &$25.12 \pm 0.10$  &$12.04 \pm 0.02$ &$ 12.22 \pm 0.03$ &$13.99 \pm 0.06$\\
i &$52.7 \pm 4.9$ &$24.96 \pm 0.10$  &$11.88 \pm 0.02$ &$ 12.05 \pm 0.03$ &$13.97 \pm 0.06$\\
\hline
\end{tabular}
\end{center}

\end{table*}

\subsection{Colour gradient of the galaxy population}
Finally, the distribution of the $g-r$ colours of the sequence
galaxies is analysed as a function of their projected distance
from the centre of the cluster. We find a significant correlation
relative to the population of faint galaxies.

\begin{figure} [!hbt]
\begin{tabular}{c}
{\psfig{figure=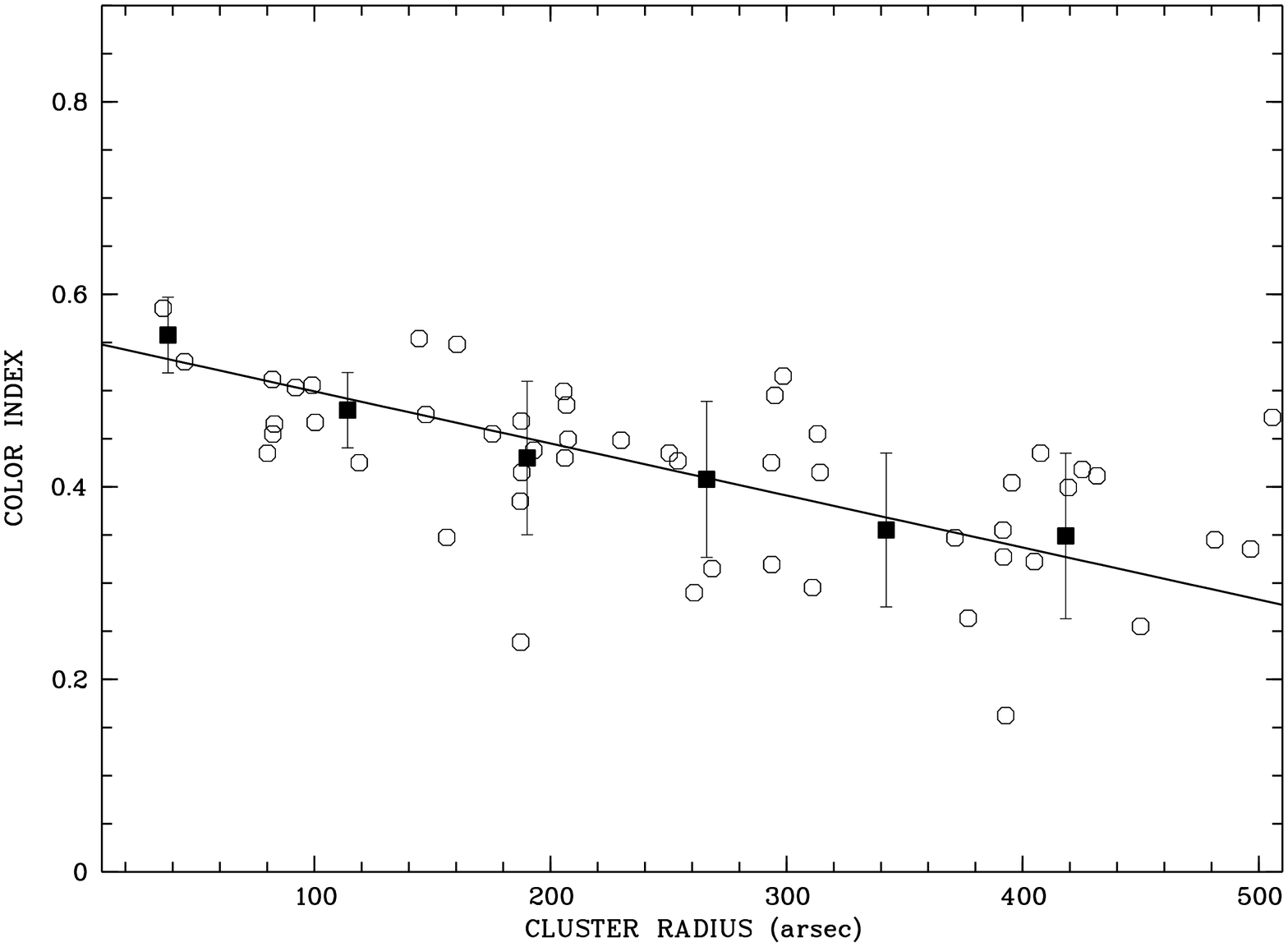,width=9cm,angle=270}}\\
{\psfig{figure=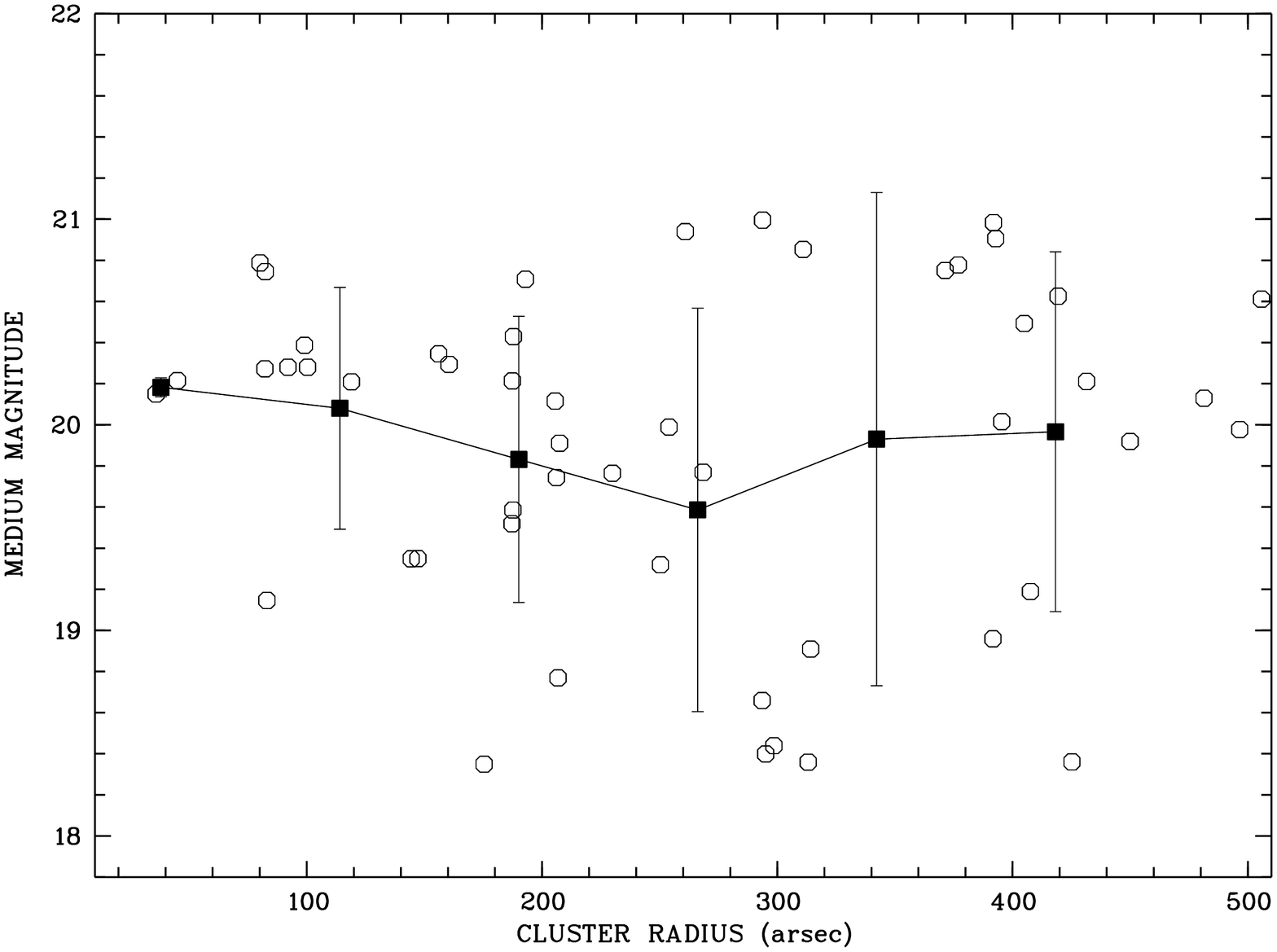,width=9cm,angle=270}}\\
\end{tabular}
\caption[]{The average colour index of dwarf sequence galaxies shows a
gradient from red to blue going off the centre of the cluster
(upper panel). This feature cannot be ascribed to the luminosity+colour
segregation: the non correlation between radius cluster and
medium magnitude of the dwarf sequence galaxies is shown (lower panel).}
\label{fig:gradi}
\end{figure}

As partly expected, brighter galaxies tend to dominate in the
central region of the cluster. Such galaxies (see also the
discussion on the $CMR$ relation) tend to be somewhat redder.
Therefore we expect a mild correlation between the cluster
integrated colour - defined as the mean colour derived from the
galaxy population located at a given distance from the centre -
and the distance from the centre. The total gradient expected to
be $< 0.2 $ in $g-r$. On the other hand if we limit ourselves to
consider only the dwarf galaxies (bottom of Fig.\ref{fig:gradi}),
we do not measure any correlation between the mean galaxy
magnitude and the distance from the cluster centre. In spite of
this lack of correlation the faint cluster population shows a
well-defined colour gradient moving outward from the centre (upper
panel of Fig. \ref{fig:gradi}). This effect is significant at a 4
sigma level and unrelated to the $CMR$ relation. Indeed over the
small range of magnitude we took into consideration ($18 <r<21$)
such an effect would be at most of about 0.1 mag, while we observe
a gradient of about 0.3 magnitudes.

A very similar result is found by Secker (1996) in Coma cluster;
conversely, Hilker et al. (1998) do not find any correlation
between the projected distance from the centre and the colours of
dwarf galaxies in the central region of Fornax cluster.

\section {Discussion and conclusion}
We started a project whose main goal is the determination of the
cluster Luminosity Function and its relation to the cluster
morphology and population. In this paper we describe the procedure
used for the analysis of the data using A496 as a test case. The
LF of this cluster has been published in Molinari et al. (1998).

In the present study, examination of the space distribution of the
blue galaxies reveals a density peak at about a core radius $\sim
500$ arcsec ($\sim 0.22$ Mpc) from the cluster centre. This
finding should be related to observation of a blueing of the
galaxy population, beginning from the cluster centre to its
outskirts (Fig. \ref{fig:gradi}). This phenomenon, demonstrated to
be independent from the CMR relation and luminosity segregation,
calls for physical differentiation in the galaxy stellar content.

Furthermore, we measure a rather blue cD halo with a remarkable
North-South colour asymmetry. This is different from what has been
found, e.g., by Molinari et al. (1994) who ascribed  the very red
cD halo was to a M0-like stellar population, implying that any
model deserves further consideration.

Further work is planned to look for these interesting features in
other clusters and a detailed discussion on the above results will
be given in a forthcoming paper of this series.

Thank are due to K. Sheldt and L. Moretti for
some help with the English proofing.


\end{document}